\documentclass[aps,prb,twocolumn,preprintnumbers,amsmath,amssymb,superscriptaddress]{revtex4}%

\usepackage{graphicx}%
\usepackage{dcolumn}
\usepackage{amsmath}
\usepackage{color}
\usepackage{multirow}

\makeatletter
\def\btt#1{\texttt{\@backslashchar#1}}%
\DeclareRobustCommand\bblash{\btt{\@backslashchar}}%
\makeatother

\begin{document}


\title{Muon spin rotation study of type-I superconductivity: \\ elemental $\beta-$Sn}

\author{Richard Karl}
 \affiliation{ETH Zurich, CH-8093 Zurich, Switzerland}
\author{Florence Burri}
 \affiliation{ETH Zurich, CH-8093 Zurich, Switzerland}
\author{Alex Amato}
 \affiliation{Laboratory for Muon Spin Spectroscopy, Paul Scherrer Institute, CH-5232 Villigen PSI, Switzerland}

\author{Mauro Doneg\`{a}}
 \affiliation{Institute for Particle Physics and Astrophysics, ETH Zurich, CH-8093 Zurich, Switzerland}
\author{Severian~Gvasaliya}
 \affiliation{Laboratory for Solid State Physics, ETH Zurich, CH-8093 Zurich, Switzerland}
\author{Hubertus Luetkens}
 \affiliation{Laboratory for Muon Spin Spectroscopy, Paul Scherrer Institute, CH-5232 Villigen PSI, Switzerland}
\author{Elvezio Morenzoni}
 \affiliation{Laboratory for Muon Spin Spectroscopy, Paul Scherrer Institute, CH-5232 Villigen PSI, Switzerland}
\author{Rustem Khasanov}
 \email{rustem.khasanov@psi.ch}
 \affiliation{Laboratory for Muon Spin Spectroscopy, Paul Scherrer Institute, CH-5232 Villigen PSI, Switzerland}

\begin{abstract}
The application of the muon-spin rotation/relaxation ($\mu$SR) technique for studying type-I superconductivity is discussed. In the intermediate state, {\it i.e.} when a type-I superconducting sample with non-zero demagnetization factor $N$ is separated into normal state and Meissner state (superconducting) domains, the $\mu$SR technique allows to determine with very high precision the value of the thermodynamic critical field $B_{\rm c}$, as well as the volume of the sample in the normal and the superconducting state.
Due to the microscopic nature of $\mu$SR technique, the $B_{\rm c}$ values are determined {\it directly} via measurements of the internal field inside the normal state domains. No assumptions or introduction of any type of measurement criteria are needed.

Experiments performed on a 'classical' type-I superconductor, a cylindrically shaped $\beta-$Sn sample, allowed to reconstruct the full $B-T$ phase diagram. The zero-temperature value of the thermodynamic critical field $B_{\rm c}(0)=30.578(6)$~mT and the transition temperature $T_{\rm c}=3.717(3)$~K were determined and found to be in a good agreement with the literature data. An experimentally obtained demagnetization factor is in very good agreement with theoretical calculations of the demagnetization factor of a finite cylinder. The analysis of $B_{\rm c}(T)$ dependence within the framework of the phenomenological $\alpha-$model allow to obtain the value of the superconducting energy gap $\Delta=0.59(1)$~meV, of the electronic specific heat $\gamma_e=1.781(3)$~${\rm mJ}/{\rm mol}\; {\rm K}^2$ and of the jump in the heat capacity ${\Delta C(T_c)}/{\gamma T_{\rm c}}=1.55(2)$.
\end{abstract}
\maketitle


\section{\label{sec:Introduction}Introduction}

Muon-spin rotation/relaxation ($\mu$SR) is a fast developing and widely used technique which is extremely sensitive to various types of magnetism.\cite{Schenk_Book_1985, Cox_JPC_1987, Dalmas_JPCM_9_1997, Yaouanc_book_2011, Blundell_ConPhys_1999, Sonier_RMP_2000, Uemura_book_2015} Compared to most other microscopic techniques by nuclear probes, a $\mu$SR experiment is relatively straightforward since the sample does not need to contain any specific nuclei. In addition, a relatively complicated sample environment can be used as illustrated, {\it e.g.} by the large number of measurements performed  down to 10~mK in temperature, up to 9.5~T in magnetic field, and up to 2.8~GPa under  pressure {\it etc.} (see {\it e.g.} Refs.~\onlinecite{Schenk_Book_1985, Cox_JPC_1987, Dalmas_JPCM_9_1997, Yaouanc_book_2011, Blundell_ConPhys_1999, Sonier_RMP_2000, Uemura_book_2015, Khasanov_PRB_2008_MoSb, Khasanov_HPR_2016, Khasanov_PRB_2018_Bi-III, Khasanov_PRB_2018_FeSe, Grinenko_PRB_2018} and references therein).

A positive muon, which is a spin-$1/2$ particle, stops in a specific place within the crystal lattice. The spin of the muon precesses around the local field $B_{\rm loc}$ at the stopping position. The precession frequency is directly related to  $B_{\rm loc}$ via the muon gyromagnetic ratio $\gamma_\mu = 2\pi\times135.5342$~MHz/T as:
\begin{equation}
\omega_\mu = \gamma_\mu B_{\rm loc}.
 \label{eq:B_loc}
\end{equation}

In the case of magnets, the field at the muon stopping site is determined by the surrounding magnetic moments (electronic and/or nuclear in origin) and by the spin density at the muon site. Since the periodicity of the magnetic structure follows, in general, the crystallographic one, resolving the type of the magnetic moment arrangement requires the knowledge of the muon stopping position, which is by itself not a trivial task.\cite{Schenk_Book_1985, Yaouanc_book_2011, Maeter_PRB_2009, Bendele_PRB_2012, Moller_PRB_2013, Moller_PhyScr_2013, Amato_PRB_2014, Mallett_EPL_2015, Khasanov_PRB_MnP_2016, Bonfa_JPSJ_2016, Khasanov_JPCM_2017, Onuorah_PRB_2018, Khasanov_PRB_2017_FeSe, Liborio_JCP_2018}
In the case of type-I and type-II superconductors the knowledge of the muon stopping site is not required. The reason is that the periodicity of various structures arising in superconductors under the applied magnetic field (as {\it e.g.} the flux-line lattice, the Meissner state, as well as various types of coexistence states) do not match the periodicity of the crystal lattice.

The character of the magnetic response, probed by means of $\mu$SR in superconducting materials, depends on the value of the externally applied field ($B_{\rm ap}$) and on the type of superconductivity.  In the Meissner state, the magnetic field is expelled completely from the sample volume except from a thin layer near the surface, where it decreases exponentially on a characteristic distance $\lambda$ ($\lambda$ is the magnetic penetration depth).\cite{Tinkham_75} The Meissner state is formed in a magnetic field lower than the first critical field ($B_{\rm c1}$) for the type-II superconductor and the thermodynamic critical field ($B_{\rm c}$) for the type-I superconductor, respectively. The exponential field decrease in the Meissner state was directly probed by muons with low and tunable energy by means of Low-Energy $\mu$SR.\cite{Morenzoni2004, Jackson_PRL_2000, Khasanov_PRL_2004, Suter_PRL_2004, Kiefl_PRB_2010}

The vortex  state forms in type-II superconductors in magnetic fields exceeding $B_{\rm c1}$. In such a case, the field penetrates the superconductor in the form of quantized flux lines (vortices) forming a regular flux-line lattice (FLL).\cite{Abrikosov_JETP_1957} Considering  the FLL to be a quasi two-dimensional object (vortices, in general, are aligned along the applied field), $\mu$SR experiments in the vortex state are performed by using muons with constant implantation energy (see {\it e.g.} Refs.~\onlinecite{Yaouanc_book_2011, Blundell_ConPhys_1999, Sonier_RMP_2000} for a review, and Refs.~\onlinecite{Schwarz_HypInt_1986, Herlach_HypInt_1990} reporting the first FLL measurements by means of $\mu$SR)

\begin{figure}[htb]
\includegraphics[width=1.03\linewidth]{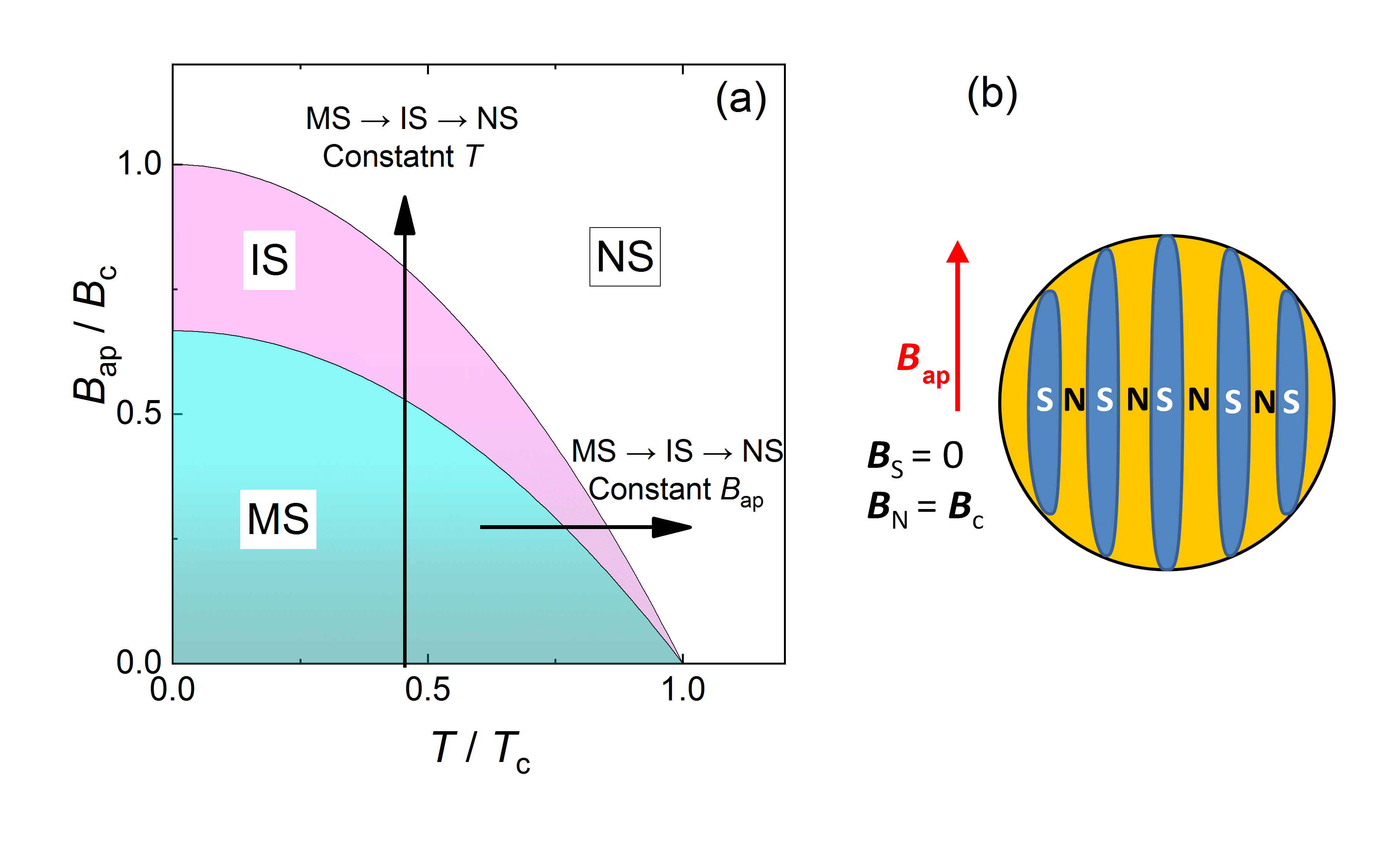}
\caption{(a) The $B-T$ phase diagram of a superconducting type-I sphere ($N=1/3$). The 'MS', 'IS' and 'NS' regions denote the parts of the phase diagram corresponding to the Meissner state (blue area), Intermediate state (pink area) and the Normal state (white area), respectively. The intermediate state develops in the region of applied fields $(1-N) B_{\rm c} < B_{\rm ap} < B_{\rm c}$.\cite{Tinkham_75, Poole_Book_2014} The vertical and horizontal arrows correspond to MS$\rightarrow$IS$\rightarrow$NS pathes at constant temperature and constant applied field $B_{\rm ap}$, respectively. (b) The separation of the superconducting type-I sphere in normal state (N) and superconducting (S) domains (after Ref.~\onlinecite{Shapiro_Arxiv_2018}). The field in the superconducting domains is equal to zero ($B_{\rm S}=0$). The field in the normal state domains is equal to the thermodynamic critical field ($B_{\rm N}= B_{\rm c}$). }
 \label{fig:Type-I}
\end{figure}

In addition to the pure Meissner and FLL states, the combination of them can be formed in superconductors with a non-zero demagnetization factor $N$. With the applied field $(1-N)B_{\rm c1} < B_{\rm ap} < B_{\rm c1}$, a type-II superconducting sample separates on the Meissner state and FLL domains, thus leading to the formation of the so called 'intermediate-mixed' state.\cite{Tinkham_75}
A type-I superconductors forms the so called 'intermediate state' by splitting itself into non-superconducting (normal state) and superconducting (Meissner state) domains for fields in the range:\cite{Tinkham_75, Poole_Book_2014}
\begin{equation}
(1-N)B_{\rm c} < B_{\rm ap} < B_{\rm c}.
 \label{eq:Intermediate-state_Interval}
\end{equation}
As an example, Figure~\ref{fig:Type-I}~a shows the $B-T$ phase diagram of a spherical type-I superconducing sample ($N=1/3$).  Figure~\ref{fig:Type-I}~b gives the schematic representation of the separation of the sample in normal state (N) and superconducting (S) domains.  Note that for bulk samples (with linear dimensions much bigger than the coherence length $\xi$) and for fields not too close to $B_{\rm c}$ and $(1-N) B_{\rm c}$ the field inside the normal state domains is practically {\it equal} to the thermodynamic critical field:\cite{Tinkham_75, Poole_Book_2014, Egorov_PRB_2001}
\begin{equation}
B_{\rm N} = B_{\rm c}
 \label{eq:B_N}
\end{equation}

So far, the vast majority of $\mu$SR experiments on superconducting materials were performed on type-II superconductors in the vortex state (see {\it e.g.} Refs.~\onlinecite{Schenk_Book_1985, Dalmas_JPCM_9_1997, Yaouanc_book_2011, Blundell_ConPhys_1999, Sonier_RMP_2000, Uemura_book_2015, Khasanov_PRB_2008_MoSb, Khasanov_HPR_2016, Khasanov_PRB_2018_Bi-III, Khasanov_PRB_2018_FeSe, Schwarz_HypInt_1986, Herlach_HypInt_1990} and references therein).
Much less work was devoted to $\mu$SR studies of the Meissner state in type-I and type-II superconductors.\cite{Jackson_PRL_2000, Khasanov_PRL_2004, Suter_PRL_2004, Kiefl_PRB_2010} Very few studies were made for type-I superconductors in the intermediate state,\cite{Gladisch_HypInt_1979, Grebinnik_JETP_1980, Egorov_PhysB_2000, Egorov_PRB_2001, Kozhevnikov_Arxiv_2018, Singh_Arxiv_2019, Beare_Arxiv_2019, Khasanov_Arxiv_2018} and, to the best of our knowledge, no $\mu$SR experiments in the intermediate-mixed state of type-II superconductors were reported so far.
The present paper discusses the application of the muon-spin rotation/relaxation  technique for studying type-I superconductors in the intermediate state, {\it i.e.} when the sample with a non-zero demagnetization factor $N$ is separated into the normal state (nonsuperconducting) and the Meissner state (superconducting) domains. We show, that due to its microscopic nature, the $\mu$SR technique allows to determine  precisely the value of the thermodynamic critical field $B_{\rm c}$ as well as the volume of the sample remaining in the normal and the superconducting (Meissner) state. In order to check the capabilities of the technique, a full $B-T$ phase diagram of a  'classical' type-I superconductor, a cylindrically shaped $\beta-$Sn sample, is reconstructed.

The paper is organized as follows. Section~\ref{sec:muSR-technique} gives the description of the transverse-field $\mu$SR technique. In Section~\ref{sec:Experiment} the experimental setup (Sec.~\ref{sec:muSR_Experiments}), the sample (Sec.~\ref{sec:Sample}),   different types of measurement approaches (field-scans, temperature-scans and $B-T$-scan, Sec.~\ref{sec:Measurement-procedure}) and  the data analysis procedure (Sec.~\ref{sec:Data-Analysis-procedure}) are discussed. Section~\ref{sec:Experimental-Results} presents the results obtained in field-scan (Sec.~\ref{sec:B-scans}), temperature-scan (Sec.~\ref{sec:T-scans}) and $B-T$-scan (Sec.~\ref{sec:B-T-scan}) sets of experiments. Effects of the magnetic history on the internal field distribution are discussed in Sec.~\ref{sec:History_Effects}. The discussion of the experimental data is given in Sec.~\ref{sec:Discussions}: the \ref{sec:Demagnetization} part of this section is devoted to the determination of the demagnetization factor $N$;  Sec.~\ref{sec:thermodynamic-field} describes the analysis of the $B_{\rm c}(T)$ dependence within the framework of the phenomenological $\alpha-$model; the comparison of physical quantities obtained in the present study with those reported in the literature is given in Sec.~\ref{sec:quantities_comparison}. Conclusions follow in Sec.~\ref{sec:Conclusions}.

\section{\label{sec:muSR-technique}Transverse-field $\mu$SR technique}

\begin{figure}[htb]
\centering
\includegraphics[width=1.0\linewidth]{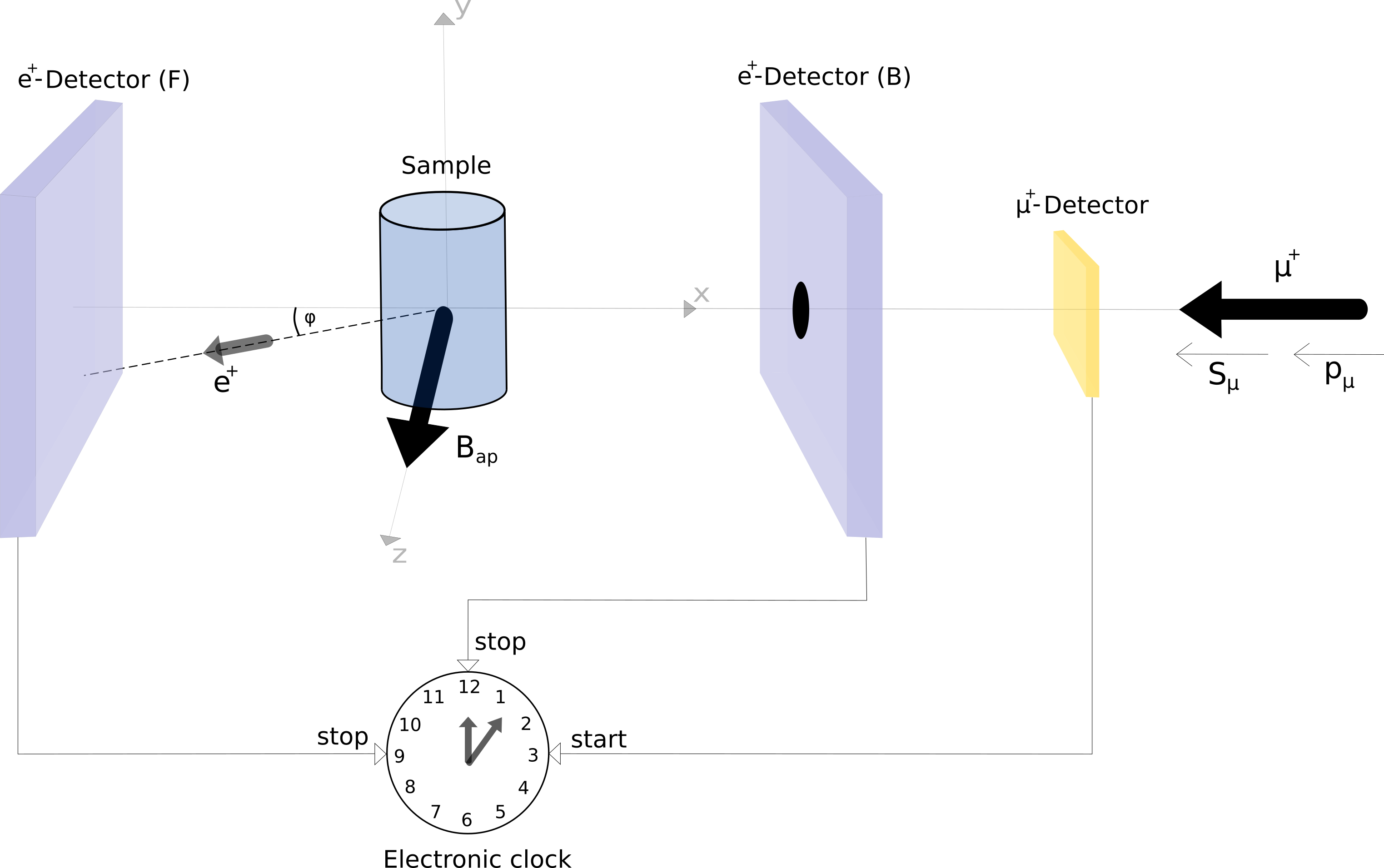}
\caption{Schematic representation of a TF-$\mu$SR experiment. Spin polarized muons with spin $\vec{S}_{\mu}$ parallel to the momentum $\vec{p}_{\mu}$ are implanted  into the sample placed between forward (F) and backward (B) positron detectors. The magnetic field $B_{\rm ap}$ is applied perpendicular to the initial muon-spin polarization.  A clock is started at the time the muon enters the start detector ($\mu^+$ detector) and stopped as soon as the decay positron is detected in one of the positron ($e^+$) detectors. Note that in the present setup, the initial muon spin direction is parallel to the momentum as we take here the so-called `Backward' muons arising from pion decaying in flight.\cite{Khasanov_HPR_2016} }
\label{fig:muSR_setup}
\end{figure}

\begin{figure}[htb]
\centering
\includegraphics[width=0.9\linewidth]{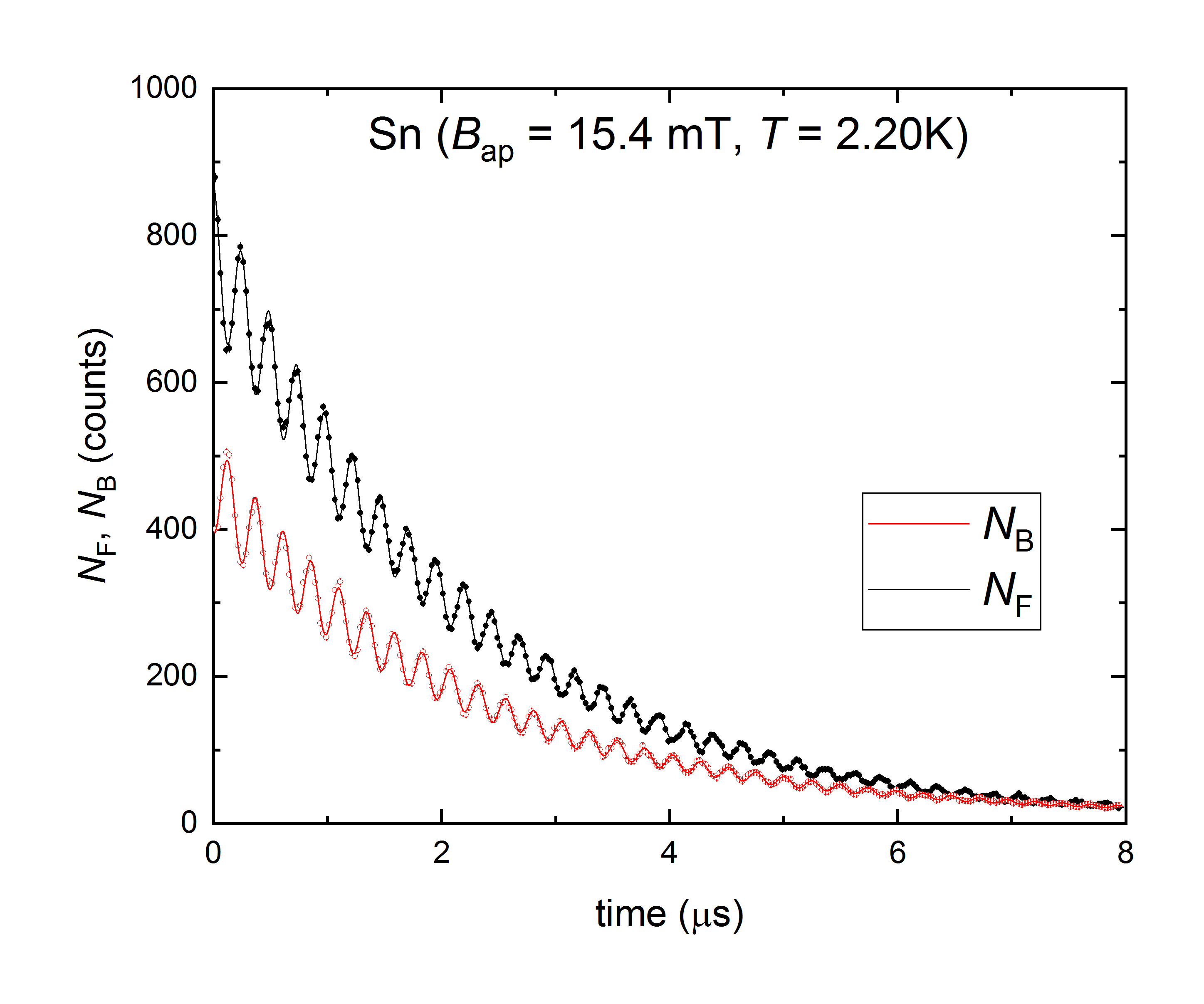}
\caption{Positron counts $N_{F}$ and $N_{B}$ as a function of time for the forward (F) and backward (B) detector, respectively. At $B_{\rm ap}=15.4$ mT and $T=2.20$ K a local magnetic field inside the Sn sample causes an oscillatory time evolution of the muon spin polarization. }
\label{fig:Histogram}
\end{figure}

The $\mu$SR method is based on the observation of the time evolution of the muon spin polarization $P(t)$ of muons that are implanted into a sample. Basic principles of transverse-field (TF) $\mu$SR experiment are illustrated in Fig.~\ref{fig:muSR_setup}. A transverse-field geometry means that the magnetic field is applied to the sample perpendicularly to the initial muon-spin polarization. At the time of the muon  implantation into the sample, an electronic clock triggered by the muon detector ($\mu^+$ detector in Fig.~\ref{fig:muSR_setup}) is started. After a very rapid thermalization, the spin of the muon precesses with the Larmor frequency (Eq.~\ref{eq:B_loc}) in the local magnetic field $B_{\rm loc}$ until the muon decays with an average lifetime of $\tau_\mu\simeq2.2\,\mu$s. A positron ($e^+$) is emitted preferentially in the direction of the muon spin at the time of its decay and it is then detected by one of the positron detectors ($e^+$ detectors in Fig.~\ref{fig:muSR_setup}) which stops the clock. As a result, a histogram as a function of time is generated for the forward [$N_{\rm F}(t)$, F denotes forward with respect to the initial spin] and the backward [$N_{\rm B}(t)$] detectors:
\begin{equation}
 N_{\rm F(B)} (t)=N_0 \exp[-t/\tau_\mu]\cdot[1 +(-) A_{\rm 0}P(t)],
\end{equation}
(see Fig.~\ref{fig:Histogram}). Here $P(t)$ is the muon-spin polarization function and $A_{0}$ is the  maximum precession amplitude at $t=0$. Note here that a time-independent background is also present in the histograms presented in Fig.~\ref{fig:Histogram}.

The time evolution of the muon polarization is further obtained either by substraction of the exponential decay component due to the muon decay, or by using the so called asymmetry function:
\begin{equation}
 A(t)=A_{0}P(t)=\frac{N_{\rm F}(t)-a N_{\rm B}(t)}{N_{\rm F}(t)+a N_{\rm B}(t)}.
\label{eq_muSR_asymmetry}
\end{equation}
The parameter $a$ takes into account the different solid angles and efficiencies of the positron detectors and it is determined by a calibration experiment. The maximum precession asymmetry $A_{0}$ depends on different experimental factors, such as the detector solid angle, the detector efficiency, the absorption, and the scattering of the positrons in the material. The values of $A_{0}$ typically lie between 0.25 and 0.3.

\section{\label{sec:Experiment}Experimental part}

In this Section, the experimental setup, the various measurement approaches and  the data analysis procedure are discussed.

\subsection{\label{sec:muSR_Experiments}Muon-spin rotation experiments}

TF-$\mu$SR experiments were carried out at the $\mu$E1 beam-line by using the dedicated GPD (General Purpose Decay) spectrometer (Paul Scherrer Institute, Switzerland).\cite{Khasanov_HPR_2016} Muons with a momentum $\simeq98$~MeV/c were used. The sample was cooled down by using an Oxford Sorption Pumped $^3$He Cryostat (base temperature $\simeq0.25$~K).
The typical counting statistics was $\sim 10^{7}$ positron events for each data point. The experimental data were analyzed using the MUSRFIT package.\cite{Suter_MUSRFIT_2012}

\subsection{\label{sec:Sample}Sn sample }

A commercial Sn sample available with $99.999\%$ purity was used as a probe. The sample was a cylinder with a diameter and a height of 20 and 100 mm, respectively. In order to avoid the transformation of the Sn sample from the superconducting $\beta-$ into the non-superconducting $\alpha-$modification, the sample was cooled down quickly (within $\sim 1.5$ hour time-period) from 300~K to $\sim 10$~K.

\subsection{\label{sec:Measurement-procedure}Measurement procedure}

\begin{figure}[htb]
\centering
\includegraphics[width=1.0\linewidth]{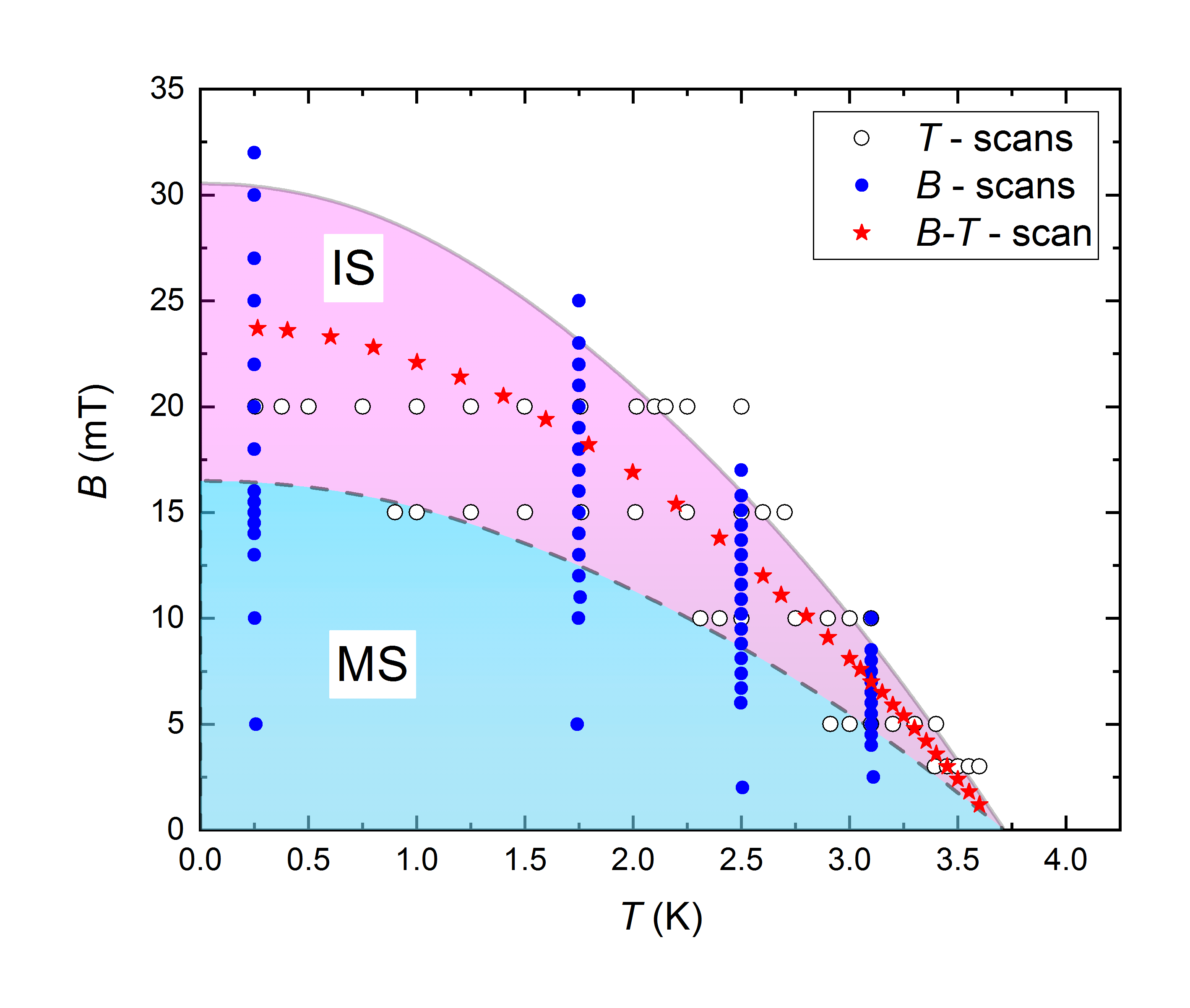}
\caption{Phase diagram with the three measurement procedures used to investigate the superconducting $\beta-$Sn. Open black symbols correspond to temperature scans taken at constant applied fields ($B_{\rm ap}=3.0$, 5.0, 10.0, 15.0 and 20.0~mT). Closed blue dots correspond to field scans taken at constant temperatures ($T=0.25$, 1.75, 2.50, and 3.10~K).  Red stars correspond to the case when both the applied field $B_{\rm ap}$ and the temperature $T$ were changed simultaneously. The intermediate state (IS) and the Meissner state (MS) areas are denoted by the pink and the blue color, respectively. The $B-T$ phase diagram of the $\beta-$Sn sample was determined for a zero-temperature value of the thermodynamic critical field $B_{\rm c}(0)=30.6$~mT, a transition temperature $T_{\rm c}=3,72$~K and a demagnetization factor $N=0.46$ (see Sec.~\ref{sec:Discussions}), assuming that the temperature evolution of $B_{\rm c}$ follows $B_{\rm c}(T)=B_{\rm c}(0)[1-(T/T_{\rm c})^2]$.\cite{Tinkham_75, Poole_Book_2014}}
\label{fig:scans}
\end{figure}

Three measurement schemes were implemented to investigate the $\beta-$Sn sample (see Fig.~\ref{fig:scans}). The first scheme ($T$-scans) involved the application of a fixed magnetic field to the sample while scanning different temperatures (corresponding to the 'constant $B_{\rm ap}$' path in Fig.~\ref{fig:Type-I}~a). The starting $B-T$ phase point was approached by a field-cooling procedure from $T$ above $T_{\rm c}$ ($T\gtrsim 4$~K). Measurements  were performed at $B_{\rm ap}=3.0$, 5.0, 10.0, 15.0 and 20.0~mT by raising the temperature and they are denoted by open points in Fig.~\ref{fig:scans}. The second scheme ($B$-scans) followed a 'constant $T$' approach shown in Fig.~\ref{fig:Type-I}~a. The sample was first cooled down in zero applied field. After adjusting the temperature, the magnetic field was increased and $\mu$SR measurements were performed following the blue points shown in Fig.~\ref{fig:scans}. The  $B$-scans were performed at $T=0.25$, 1.75, 2.50, and 3.10~K. The third scheme ($B-T$-scan) corresponds to the case when both the applied field and the temperature were changed simultaneously. The idea was to follow the $B-T$ path of Sn's phase diagram (red stars in Fig.~\ref{fig:scans}) allowing to keep the volume parts of the sample equally occupied by the normal state and the Meissner state domains ($f_{\rm N}\simeq f_{\rm S}$, $f$ denotes the volume fraction). For doing so, the $B-T$ points were taken exactly in between the $(1-N)B_{\rm c}(T)$ and $B_{\rm c}(T)$ curves which, according to Fig.~\ref{fig:Type-I}~a and Eq.~\ref{eq:Intermediate-state_Interval}, determine the lower and the upper border of the intermediate state of a type-I superconductor.

\subsection{\label{sec:Data-Analysis-procedure}Data analysis procedure}

The experimental data were analyzed by separating the TF-$\mu$SR response of the sample (s) and the background (bg) contributions:
 \begin{equation}
A_0  P(t) = A_{\rm s} P_{\rm s}(t) + A_{\rm bg} P_{\rm bg}(t).  \label{eq:P(t)}
\end{equation}
Here $A_0$ is the initial asymmetry of the muon-spin ensemble. $A_{\rm s}$ ($A_{\rm bg}$) and $P_{\rm s}(t)$ [$P_{\rm bg}(t)$] are the asymmetry  and the time evolution of the muon spin polarization of the sample (background), respectively.  The background part accounts for the muons stopped outside the sample ({\it e.g.} in the sample holder or/and in the cryostat's radiation shields). Within the full set of experiments the background asymmetry $A_{\rm bg}$ did not exceed $\simeq 8\%$ of the initial asymmetry $A_0$.

The background contribution was described as:
 \begin{equation}
P_{\rm bg}(t) = e^{-\lambda_{\rm bg} t}\; \cos (\gamma_\mu B_{\rm ap} t +\phi), \label{eq:Background_P(t)}
\end{equation}
where $\lambda_{\rm bg}$ is the exponential relaxation rate, $B_{\rm ap}$ is the applied field, and $\phi$ is the initial phase of the muon-spin ensemble.

The sample contribution was described by assuming a separation between normal state (N) and superconducting (S) domains:
 \begin{eqnarray}
P_{\rm s}(t) &=&f_{\rm N} \; e^{-\lambda_{\rm N} t} \; \cos (\gamma_\mu B_{\rm N} t +\phi) +\nonumber\\
&&  (1-f_{\rm N})  \left[ \frac{1}{3} + \frac{2}{3}(1-\sigma_{\rm GKT}^2t^2)e^{-\sigma_{\rm GKT}^2t^2/2} \right]. \label{eq:Sample_P(t)}
\end{eqnarray}
Here, the first term in the right-hand site of the equation corresponds to the sample's normal state response: $f_{\rm N}$ is the normal state volume fraction ($f_{\rm N}=1$ for $T\geq T_{\rm c}$), $\lambda_{\rm N}$ is the exponential relaxation rate and $B_{\rm N}$ is the internal field [$B_{\rm N}=B_{\rm c}$ for $T<T_{\rm c}(B_{\rm ap})$ and $B_{\rm N}=B_{\rm ap}$  for $T\geq T_{\rm c}(B_{\rm ap})$, respectively]. The second term describes the contribution of the superconducting part of the sample remaining in the Meissner state ($B_{\rm S}=0$). It is approximated by the Gaussian Kubo-Toyabe function with the relaxation rate $\sigma_{\rm GKT}$, which is generally used to describe the nuclear magnetic moment contribution in zero-field experiments (see {\it e.g.} Ref.~\onlinecite{Yaouanc_book_2011} and references therein).

\begin{figure}[htb]
\centering
\includegraphics[width=1.0\linewidth]{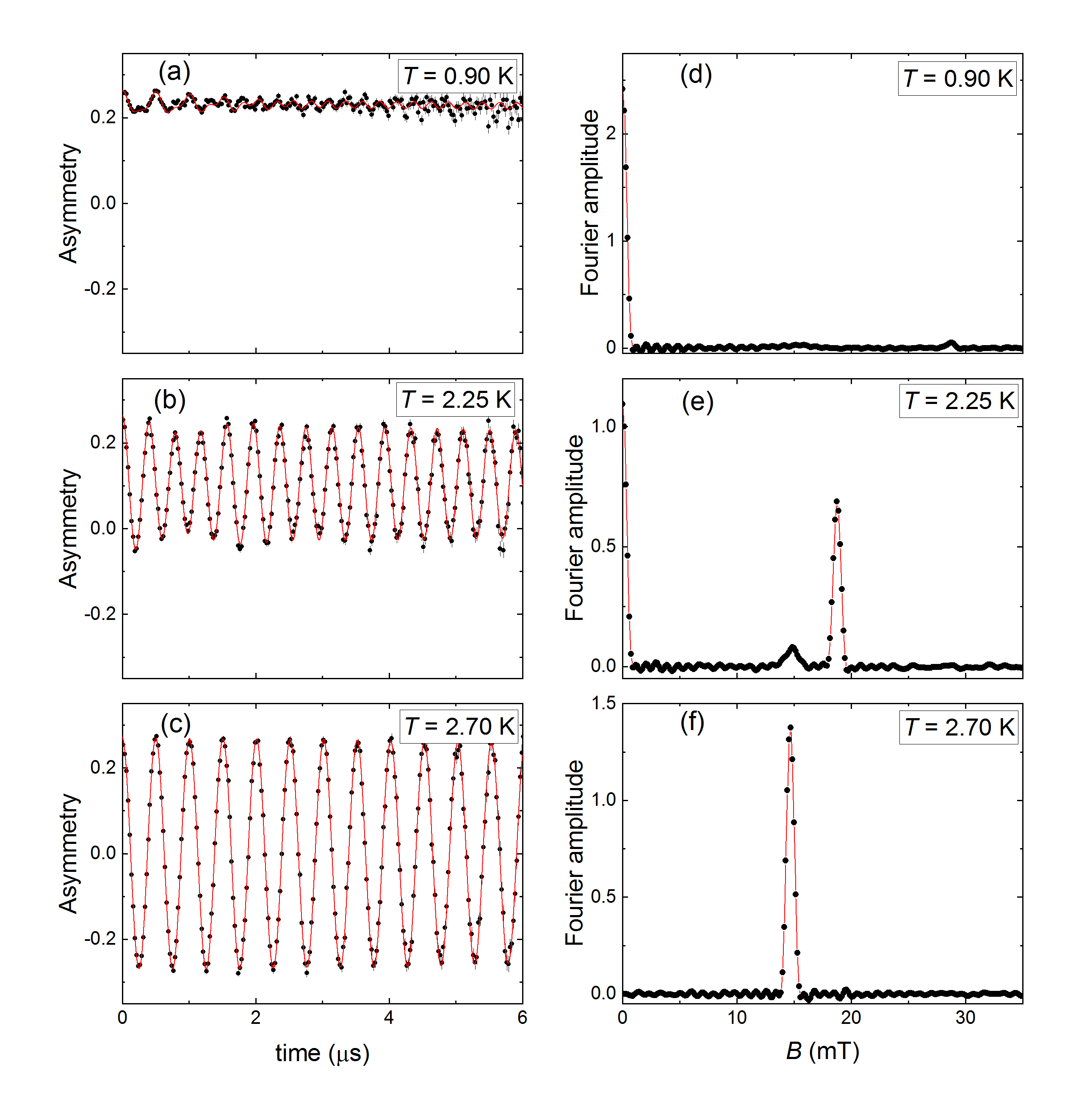}
\caption{(a) -- (c) TF-$\mu$SR time spectra taken at $B_{\rm ap}=15$~mT at $T\simeq 0.90$~K (panel a), 2.25~K (panel b)  and 2.70~K (panel c). The red lines are fits of Eq.~\ref{eq:P(t)} (with the background and the sample contributions described by Eqs.~\ref{eq:Background_P(t)} and \ref{eq:Sample_P(t)}) to the experimental data. (d) -- (e) Magnetic field distribution in the $\beta-$Sn sample obtained via Fourier transformation of the TF-$\mu$SR time spectra shown in panels (a) -- (c).}
\label{fig:Time-spectra}
\end{figure}

The fit of Eq.~\ref{eq:P(t)} to the data was performed globally. The $\mu$SR time spectra obtained within each individual $T$-scan experiment were fitted simultaneously with $A_{\rm s}$, $A_{\rm bg}$, $B_{\rm ap}$, $\sigma_{\rm GKT}$, and $\lambda_{\rm N}$ as common parameters, and $f_{\rm N}$, $B_{\rm N}$, and $\lambda_{\rm bg}$ as individual parameters for each particular data set. In the $B$- and $B-T$-scan experiments $A_{\rm s}$, $A_{\rm bg}$, $\sigma_{\rm GKT}$, and $\lambda_{\rm N}$ were fitted globally, and $B_{\rm ap}$, $f_{\rm N}$, $B_{\rm N}$, and $\lambda_{\rm bg}$ were fitted individually for each particular point.

Figures \ref{fig:Time-spectra}~a--c show TF-$\mu$SR time spectra taken at $T\simeq 0.90$~K (panel a), 2.25~K (panel b) and 2.70~K (panel c)  at an applied magnetic field $B_{\rm ap}=15$~mT. The solid red lines are fits of Eq.~\ref{eq:P(t)} to the asymmetry spectra, with the background and the sample contributions described by Eqs.~\ref{eq:Background_P(t)} and \ref{eq:Sample_P(t)}.  The corresponding magnetic field distributions obtained via Fourier transformation of TF-$\mu$SR time spectra are presented in Figs.~\ref{fig:Time-spectra} d--f.

Obviously, the behavior observed at $T=0.90$~K (panels a and d of Fig.~\ref{fig:Time-spectra}) corresponds to the response of the $\beta-$Sn sample remaining in the Meissner state (see also Fig.~\ref{fig:scans}). Indeed, the field inside the vast majority of the sample volume is equal to zero, while only a small amount of the sample remains in the intermediate state (see the high-intensity sharp peak at $B=0$ and the weak peak at $B\simeq 29$~mT, Fig.~\ref{fig:Time-spectra}~d). Fit of Eq.~\ref{eq:P(t)} to the data results in $f_{\rm N}\simeq 0.0346(2)$ and $B_{\rm N}=28.69(4)$~mT. At $T=2.25$~K (panels b and e of Fig.~\ref{fig:Time-spectra}) the Sn sample is clearly separated into the normal state and the superconducting domains. The fit results in $f_{\rm N}= 0.562(2)$, thus suggesting that the normal state domains occupy more than half of the sample volume, and $B_{\rm N}=B_{\rm c}=18.76(2)$~mT. At $T=2.70$~K (panels c and f of Fig.~\ref{fig:Time-spectra}) the field inside the sample coincides with the applied field $B_{\rm N}=B_{\rm ap}$ and no peak at $B=0$ is anymore present. This indicates that at $T=2.70$~K and $B_{\rm ap}\simeq 15$~mT the Sn sample is already in the normal state (see also Fig.~\ref{fig:scans}).

At the end of this Section we would mention that the $\mu$SR technique has no spatial resolution, so the exact domain structure in the intermediate state of type-I superconductor cannot be resolved. Only the 'integrated' quantities, as the value of fields inside the normal state and the superconducting domains, as well as relative volume fractions of various domains can be obtained.

\section{\label{sec:Experimental-Results}Experimental results}

This Section presents the experimental data obtained by following the three measurement procedures described previously.

\subsection{\label{sec:B-scans}Field scans}

In this set of experiments the applied field $B_{\rm ap}$ was scanned while the temperature was kept fixed. Experiments were performed at $T=0.25$, 1.75, 2.50, and 3.10~K. The measurement points are denoted by blue closed dots in Fig.~\ref{fig:scans}.

\begin{figure}[htb]
\includegraphics[width=1.1\linewidth]{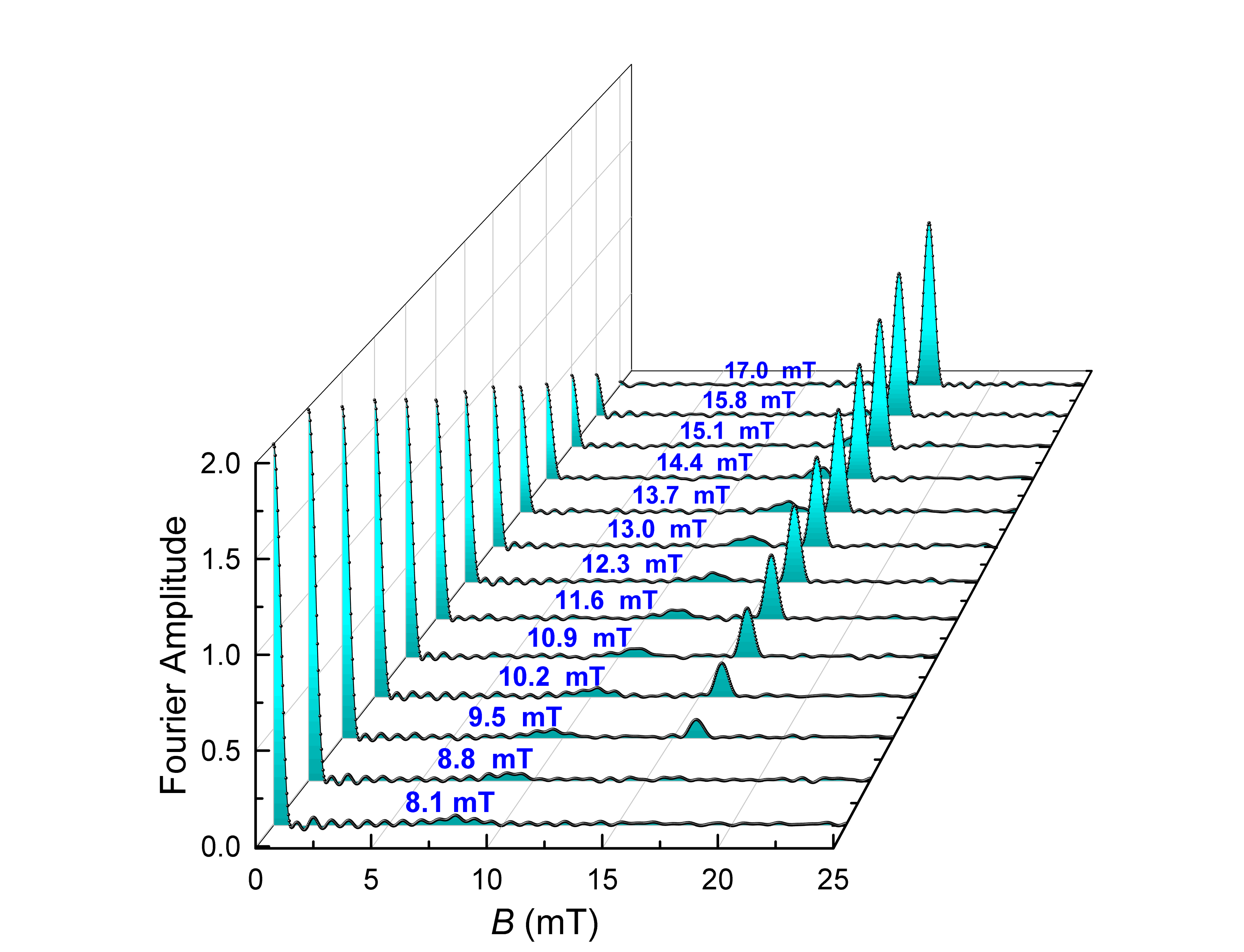}
\caption{Magnetic field distribution in the $\beta-$Sn sample at various magnetic fields and $T=2.50$~K. In the intermediate state ($9.0{\rm ~mT}< B_{\rm ap} < 16.0$~mT) two peaks at $B=0$ and $B_{\rm N}=B_{\rm c}>B_{\rm ap}$ are present. The broad peak at the applied field position corresponds to the background contribution.}
\label{fig:B-scans_Fourier}
\end{figure}

The magnetic field distributions in the $\beta-$Sn sample measured at $T=2.50$~K are shown in Fig.~\ref{fig:B-scans_Fourier}. For fields $B_{\rm ap} \lesssim 9$~mT the sample is in the Meissner state. The magnetic field distribution consists of a sharp peak at $B=0$ and a broad low-intensity peak at $B=B_{\rm ap}$ which is attributed to the background contribution. By further increasing the field, a clear peak at $B=B_{\rm c}>B_{\rm ap}$ appeared thus suggesting the separation of the sample on the normal state and the superconducting domains. The intensity of the $B>B_{\rm ap}$ and the $B=0$ peaks shows opposite tendency: while the intensity of the first one increases, the intensity of $B=0$ peak decreases until it disappears for fields exceeding $\simeq 16$~mT.

\begin{figure}[htb]
\centering
\includegraphics[width=0.85\linewidth]{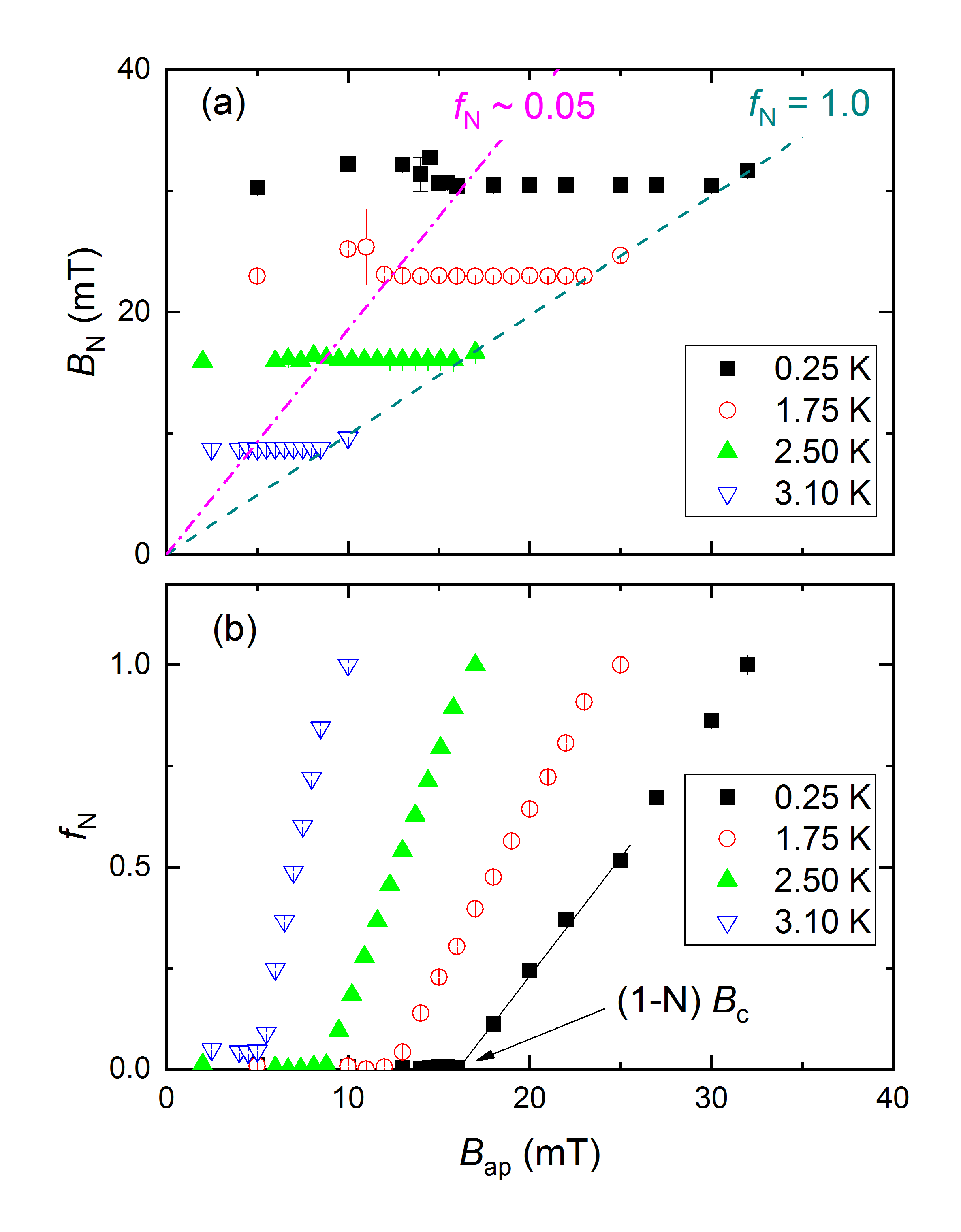}
\caption{(a) Dependence of the internal fields in the normal state domains $B_{\rm N}$ on the applied field $B_{\rm ap}$ at $T=0.25$, 1.75, 2.50 and 3.10~K. The dashed and dash-dotted lines correspond to $f_{\rm N}=1.0$ and $f_{\rm N}\sim 0.05$, respectively. (b) Dependence of the normal state volume fraction  $f_{\rm N}$ on the applied field $B_{\rm ap}$ at $T=0.25$, 1.75, 2.50 and 3.10~K. The field of $(1-N) B_{\rm c}$ is determined from the intersection of the linear fit of $f_{\rm N}(B_{\rm ap})$ in the region $0.1\lesssim f_{\rm N} \lesssim 0.5$ with the $f_{\rm N}=0$ line, see text for details.}
\label{fig:BnFn-Hscan}
\end{figure}

Figure~\ref{fig:BnFn-Hscan} shows the dependence of the fit parameters (internal field $B_{\rm N}$ and volume fraction $f_{\rm N}$ of the normal state domains) on the applied field obtained from the field-scan set of experiments. The dashed and dash-dotted lines in panel a, labelled as $f_{\rm N}=1.0$ and $f_{\rm N}\sim 0.05$, determine the region of existence of the intermediate state in the cylindrical Sn sample.
Obviously, the internal field in the normal state domains (Fig.~\ref{fig:BnFn-Hscan}~a) stays constant in the intermediate state region and follows the applied field ($B_{\rm N}=B_{\rm ap}$) as soon as $f_{\rm N}$ reaches unity ({\it i.e.} when the sample completely transforms into the normal state).  Considering that the field within the normal state domains in the intermediate state of type-I superconductor is equal to the thermodynamic critical field $B_{\rm c}$ (see Eq.~\ref{eq:B_N}), $B_{\rm c}$ at any particular temperature was obtained by averaging $B_{\rm N}$ values measured between $f_{\rm N}=1.0$ and $f_{\rm N}\sim 0.05$ curves. The analysis gives $B_{\rm c}(0.25{\rm ~K})=30.422(11)$~mT, $B_{\rm c}(1.75{\rm ~K})=22.976(7)$~mT, $B_{\rm c}(2.50{\rm ~K})=16.075(8)$~mT, and $B_{\rm c}(3.10{\rm ~K})=8.786(5)$~mT.

At the end of this Section, we want to mention an unexpected feature of field-scan results taken at $T=3.10$~K. The volume fraction of the normal state domains $f_{\rm N}$ measured at $T=3.10$~K never reaches the zero value, in contrast to the behavior observed at $T=0.25$, 1.75, and 2.50~K (see Fig.~\ref{fig:BnFn-Hscan}~b). This implies that a pure Meissner state does not set in at 3.10~K, unlike the case at lower temperatures. The reasons for such an effect are still not clear and require further investigations.

\subsection{\label{sec:T-scans}Temperature scans}

In the $T$-scan set of measurements, the temperature was scanned by keeping the applied field $B_{\rm ap}$ fixed. Experiments were performed at $B_{\rm ap}=3.0$, 5.0, 10.0, 15.0 and 20.0~mT. The measurement points are denoted by open black symbols in Fig.~\ref{fig:scans}.

\begin{figure}[htb]
\includegraphics[width=1.1\linewidth]{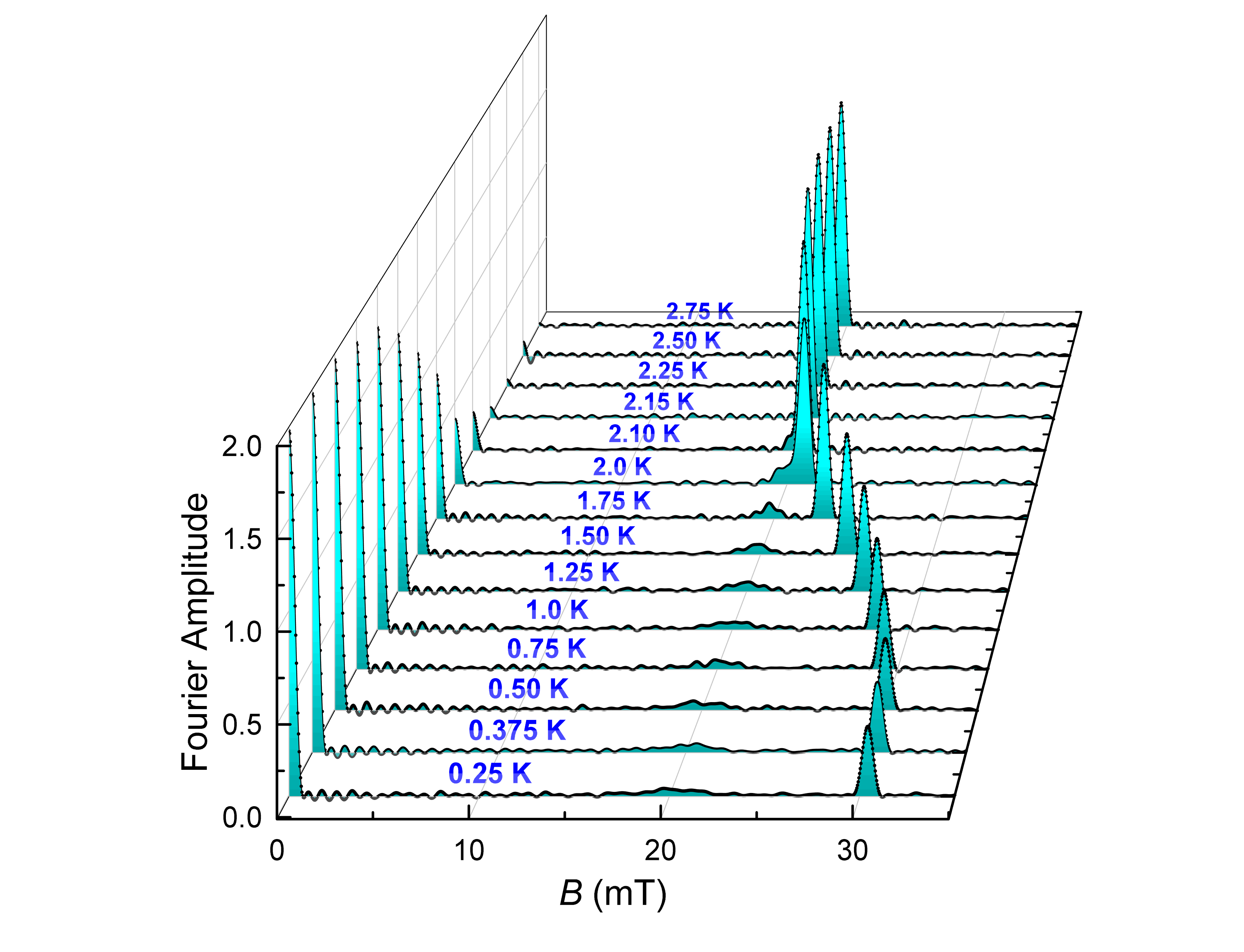}
\caption{Magnetic field distribution in the $\beta-$Sn sample at various temperatures at $B_{\rm ap}=20.0$~mT. The peak at the applied field position corresponds to the background contribution.}
\label{fig:T-scans_Fourier}
\end{figure}

The magnetic field distributions in the $\beta-$Sn sample measured at $B_{\rm ap}=20.0$~mT are presented in Fig.~\ref{fig:T-scans_Fourier}. It is obvious that even at the lowest temperature ($T=0.25$~K) the cylindrical Sn sample stays in the intermediate state. Upon increasing the temperature from 0.25 to 2.1~K, two tendencies are clearly visible: (i) The intensities of the $B>B_{\rm ap}$ and $B=0$ peaks behave in an opposite way. The increase of the $B>B_{\rm ap}$ peak intensity is reflected by a corresponding decrease of the $B=0$ peak intensity. (ii) By increasing the temperature, the $B>B_{\rm ap}$ peak shifts towards $B_{\rm ap}$. Bearing in mind that the field in the normal state domains in the intermediate state of type-I superconductors is equal to the thermodynamic critical field (see Eq.~\ref{eq:B_N}), the $T-$dependence of $B>B_{\rm ap}$ peaks represents the temperature evolution of $B_{\rm c}$. At higher temperatures only a single peak at $B=B_{\rm ap}$ is visible thus indicating that for $T\gtrsim 2.15$~K the Sn sample stays in the normal state.

\begin{figure}[htb]
\centering
\includegraphics[width=0.85\linewidth]{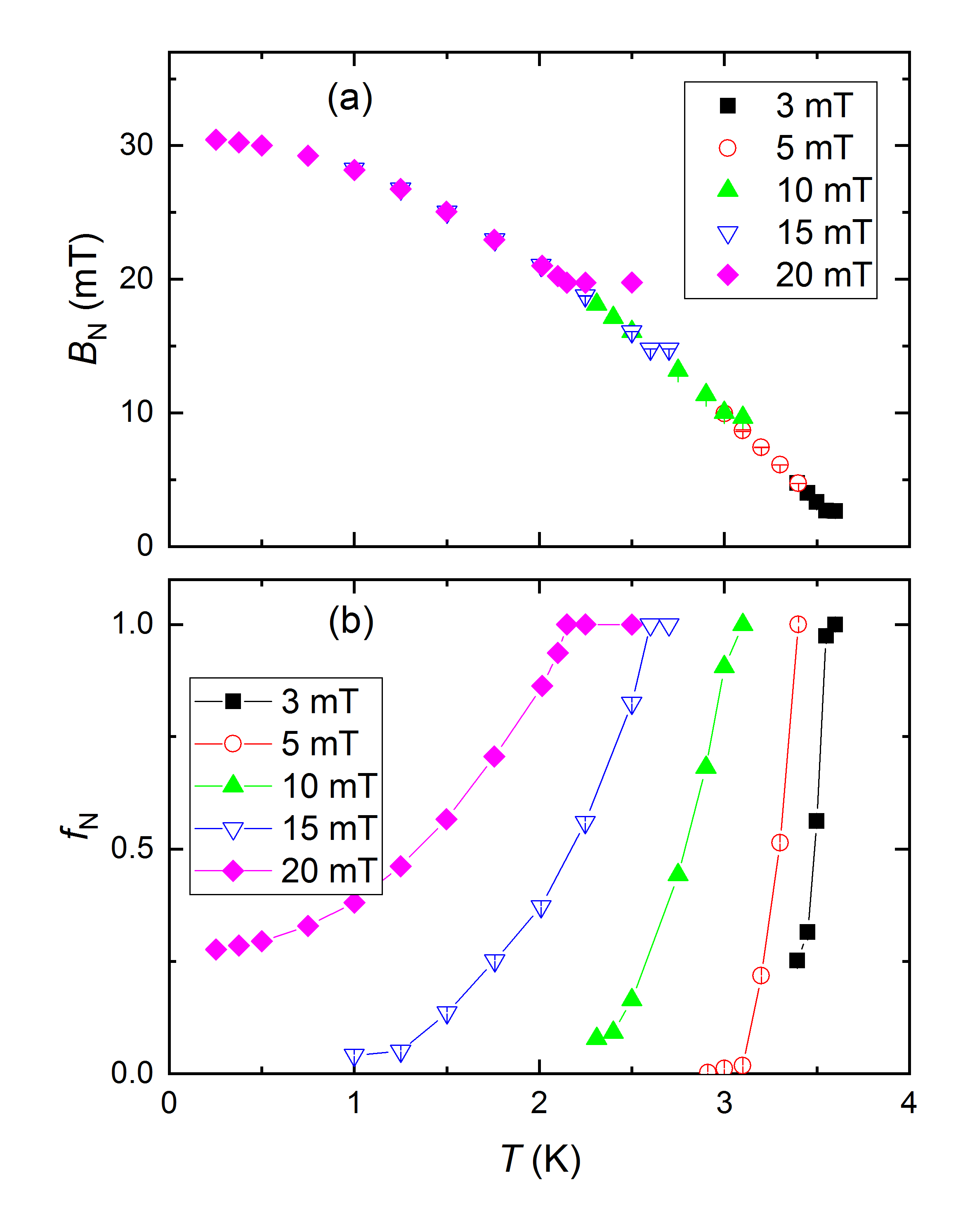}
\caption{(a) Temperature dependence of the internal field in the normal state domains $B_{\rm N}$ at $B_{\rm ap}=3.0$, 5.0, 10.0, 15.0, and 20.0~mT. (b) Dependence of the normal state domain volume fraction  $f_{\rm N}$ on temperature  at $B_{\rm ap}=3.0$, 5.0, 10.0, 15.0, and 20.0~mT.}
\label{fig:BnFn-Tscan}
\end{figure}

Figure~\ref{fig:BnFn-Tscan} shows the dependence of the fit parameters $B_{\rm N}$ and $f_{\rm N}$ on temperature. For $0<f_{\rm N}<1.0$ the dependence of $B_{\rm N}$ on $T$ reflects the temperature evolution of the thermodynamic critical field $B_{\rm c}$. In the normal state ($f_{\rm N}=1$) $B_{\rm N}$ is equal to the applied field $B_{\rm N}=B_{\rm ap}$.

It is interesting to note, that similar $B_{\rm c}(T)$ values can be  obtained at different applied fields (see the overlapping points in Fig.~\ref{fig:BnFn-Tscan}~a). The reason for such overlapping is that the intermediate state formation condition, as is defined by Eq.~\ref{eq:Intermediate-state_Interval}, is fulfilled for different $B_{\rm ap}$'s. This directly confirms that the internal field in the normal state domains of type-I superconductors in the intermediate state (at least inside bulk samples as the one studied here) is {\it independent} on the relative volumes occupied by the normal state ($f_{\rm N}$) and the superconducting ($1-f_{\rm N}$) domains. To the best of our knowledge, this is the first direct confirmation of such a statement.

The independence of $B_{\rm c}$ on $f_{\rm N}$ at constant temperature also follows from the results of the $B$-scan experiments (Sec.~\ref{sec:B-scans} and Fig.~\ref{fig:BnFn-Hscan}). All together, the results of Secs.~\ref{sec:B-scans} and \ref{sec:T-scans}  confirm the calculations presented in Refs.~\onlinecite{Tinkham_75, Egorov_PRB_2001, deGennes_Book_1966} revealing that the maximum difference between $B_{\rm N}$ and $B_{\rm c}$ (which is observed close to the Meissner state to the intermediate state transition and vanishes by approaching the normal state border) does not exceed $B_{\rm c}\cdot 2\delta_{\rm d}/d$. Here $\delta_{\rm d}=\xi-\lambda$ is the difference between coherence length $\xi$ and magnetic penetration depth $\lambda$, and $d=20$~mm is the sample diameter. For $\beta-$Sn with $\xi=230$~nm, $\lambda=34$~nm,\cite{Kittel_Book_1996} one gets $B_{\rm c}-B_{\rm N}\lesssim 3\cdot 10^{-4}$~mT, which is at least one order of magnitude smaller than the accuracy of the $B_{\rm loc}$ determination in our $\mu$SR studies (see {\it e.g.} $B_{\rm c}$'s obtained in the field-scan experiments, Sec.~\ref{sec:B-scans}).

\subsection{\label{sec:B-T-scan}$B-T$-scan}

In this set of experiments both the applied field and the temperature were changed simultaneously along the path where the volumes occupied by the normal state and the Meissner state (superconducting) domains are equal: $f_{\rm N}\simeq f_{\rm S}\sim 0.5$. The measurement points of the $B-T$-scan experiment are presented by red stars in Fig.~\ref{fig:scans}.

\begin{figure}[htb]
\centering
\includegraphics[width=0.85\linewidth]{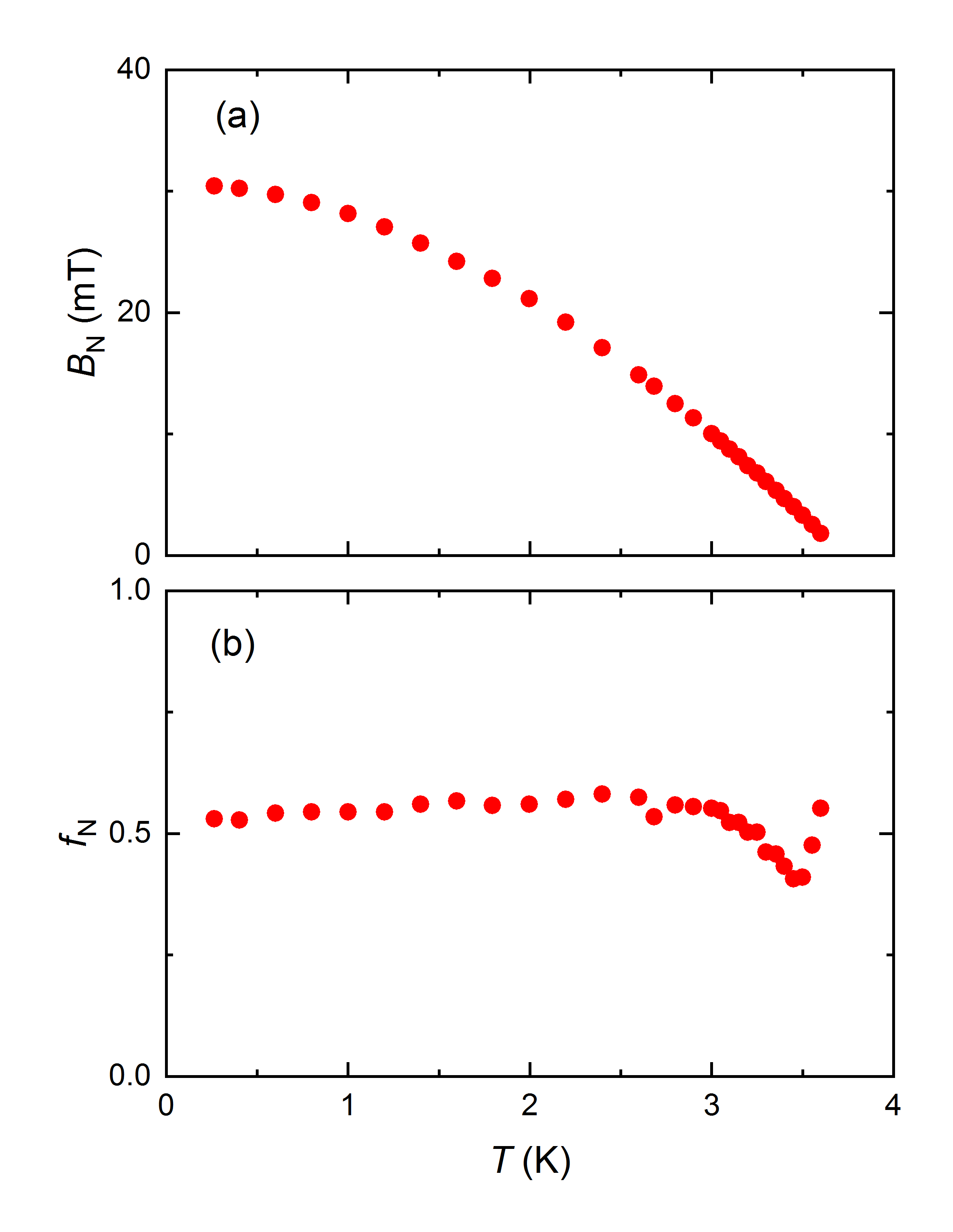}
\caption{(a) Temperature dependence of the internal field in the normal state domains $B_{\rm N}$ obtained in $B-T$-scan experiments. (b) Temperature dependence of the normal state domain volume fraction  $f_{\rm N}$ in $B-T$-scan measurments.}
\label{fig:BnFn_H-T_scan}
\end{figure}

Figure~\ref{fig:BnFn_H-T_scan} shows the dependence of the fit parameters $B_{\rm N}$ and $f_{\rm N}$ on temperature as obtained in the $B-T$-scans. Following previous discussions, $B_{\rm N}(T)$ (Fig.~\ref{fig:BnFn_H-T_scan}~a) represent the temperature evolution of the thermodynamic critical field $B_{\rm c}$.

The non-monotonic behavior of $f_{\rm N}$ (Fig.~\ref{fig:BnFn_H-T_scan}~b) requires further comments. Note that prior of the $B-T$-scan experiments, the measurement  points $(B_{\rm ap},T)$ (red stars in  Fig.~\ref{fig:scans}) were calculated to obtain equal volume fractions of the normal state and the superconducting domains ($f_{\rm N}= f_{\rm S}= 0.5$). This works, however, only up to $T\simeq 3.0$~K (see Fig.~\ref{fig:BnFn_H-T_scan}~b). Above this temperature, $f_{\rm N}$ first decreases down to $\sim0.35$ and then increases up to $f_{\rm N}\sim 0.6$ by approaching $T_{\rm c}(B=0)\simeq 3.72$~K. It is currently unclear if this effect has the similar origin as the absence of a purely Meissner state in $T=3.10$~K field-scan experiments (Sec.~\ref{sec:B-scans} and Fig.~\ref{fig:BnFn-Hscan}~b), or it stems from uncertainties of our $(B_{\rm ap},T)$ point calculations.

\subsection{\label{sec:History_Effects}Magnetic history effects}

The domain structure of type-I superconductors was found to depend on the magnetic history. In series of papers Prozorov {\it et al.}\cite{Prozorov_PRL_2007, Prozorov_NatPhys_2008, Prozorov_JPCS_2009} have shown that based on the way of how the final $B-T$ point is reached, different types of domain structures might be realized. In order to check the influence of the magnetic history on the TF-$\mu$SR response, several $B-T$ points approached by  different measurement schemes were examined.

\begin{figure}[htb]
\centering
\includegraphics[width=0.85\linewidth]{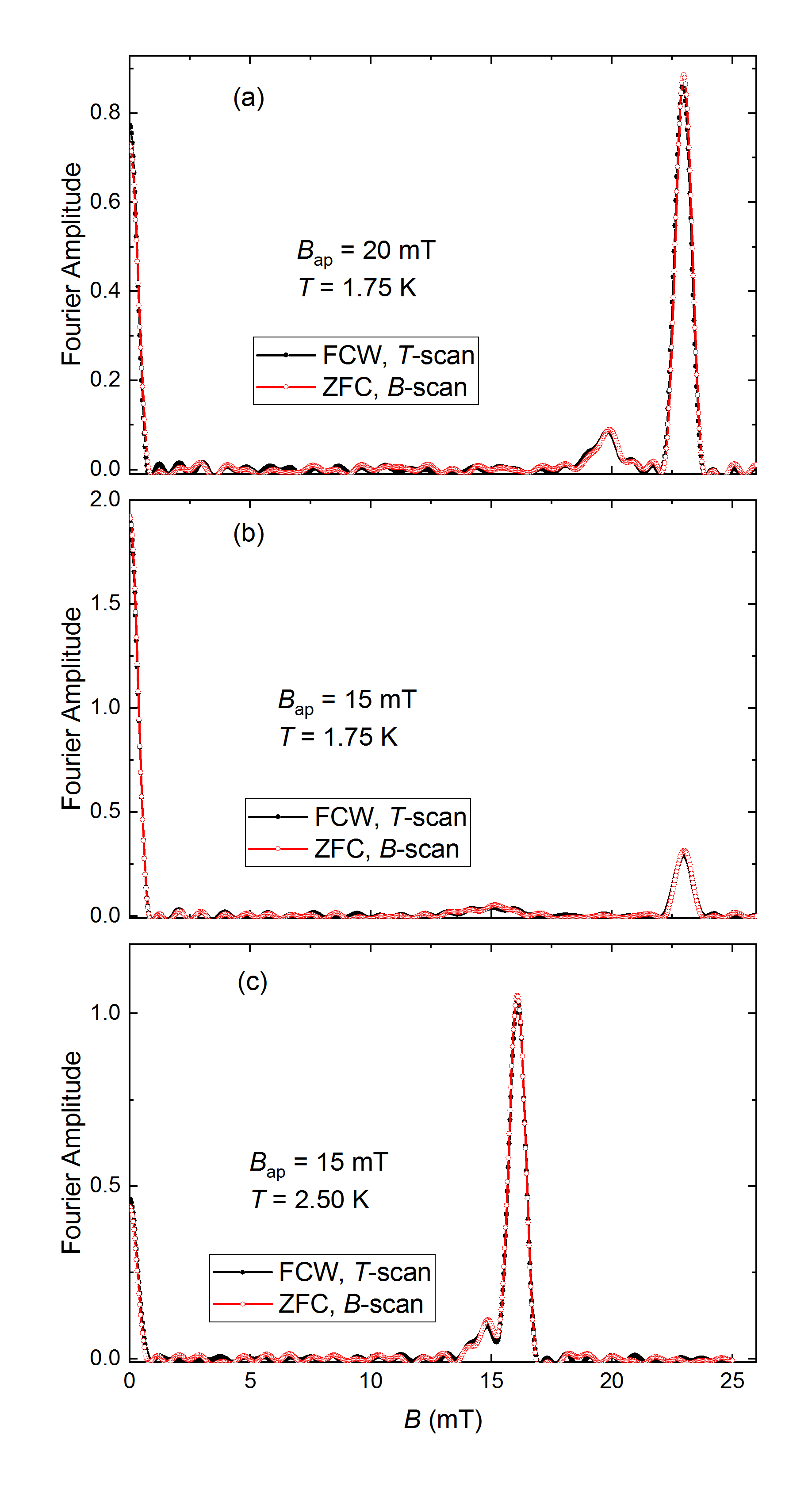}
\caption{The magnetic field distribution in a cylindrical $\beta-$Sn sample obtained in field-cooling warming (FCW, $T$-scan) and zero-field cooling (ZFC, $B$-scan) set of experiments. (a) $B_{\rm ap}=20.0$~mT, $T=1.75$~K. (b) $B_{\rm ap}=15.0$~mT, $T=1.75$~K. (c) $B_{\rm ap}=15.0$~mT, $T=2.50$~K.}
\label{fig:history}
\end{figure}

Figure \ref{fig:history} shows the distribution of magnetic fields in the cylindrical $\beta-$Sn sample obtained in $B$-scan and $T$-scan measurements. Three data sets taken at $B_{\rm ap}=20.0$~mT and $T=1.75$~K (panel a); $B_{\rm ap}=15.0$~mT, $T=1.75$~K (panel b); and $B_{\rm ap}=15.0$~mT, $T=2.50$~K (panel c) are presented. According to the measurement process description (Sec.~\ref{sec:Measurement-procedure}), in the  $B$-scan experiments, the $B-T$ points are reached by following the zero-field cooling (ZFC) path: the sample was first cooled down in zero applied field and then, by keeping the temperature constant, $B_{\rm ap}$ was continuously increased until the sample transformed into the normal state. In $T$-scan experiments the sample was first cooled down to 0.25~K in constant field and then, without changing the applied field, the temperature was increased up to the measuring one. Such process corresponds to the field-cooling warming (FCW) path.

Figure \ref{fig:history} shows that both ZFC and FCW pathes result in similar field distributions. The parameters obtained from the fit of TF-$\mu$SR spectra are found to be the same within the experimental uncertainty. This indicates that if even the distribution and/or the shape of the domains are history dependent,\cite{Prozorov_PRL_2007, Prozorov_NatPhys_2008, Prozorov_JPCS_2009} the internal field inside the normal state domains as well as the relative sample volumes occupied by the normal and the Meissner state domains remain {\it unchanged}.

\section{\label{sec:Discussions}Discussions}

In this Section the results reported in Sec.~\ref{sec:Experiment} are discussed. A particular attention is paid to the physical quantities which can be obtained from TF-$\mu$SR studies of type-I superconductor.

\subsection{\label{sec:Demagnetization}Demagnetization effects}

The intermediate state in type-I superconductors may only be formed in a sample with a non-zero demagnetization factor $N$. Following Ref.~\onlinecite{Prozorov_PRAppl_2018}, the theoretical demagnetization factor value ($N_{\rm th}$) for a finite cylinder in a magnetic field applied perpendicular to the cylinder axis is given by:
\begin{equation}
\frac{1}{N_{\rm th}} = 2+\frac{1}{\sqrt{2}}\frac{d}{l},
 \label{eq:Demagnetization-factor_theory}
\end{equation}
where $d$ and $l$ are the diameter and the length of the cylinder, respectively. For the cylindrical $\beta-$Sn sample studied here ($d=20$~mm and $l=100$~mm, Sec.~\ref{sec:Sample}) Eq.~\ref{eq:Demagnetization-factor_theory} results in $N_{\rm th}=0.467$.

The experimental value of the demagnetization factor can be estimated in two ways: from $f_{\rm N}(B_{\rm ap})$ dependencies obtained in field-scan studies and by scaling the $f_{\rm N}(B_{\rm ap})$ and $f_{\rm N}(T)$ curves as reported in the field-scan and temperature-scan experiments.

It is important to note here that the exact value of the demagnetization factor is found only for the sample of ellipsoidal shape. The cylindrically shaped  Sn sample studied here is not ellipsoidal and, strictly speaking, the demagnetization factor becomes a function of position inside the sample.\cite{Prozorov_PRAppl_2018} In such a case one should refer to the so-called {\it effective} demagnetization factor. The demagnetization factors discussed in the forthcoming Sections~\ref{sec:Demagnetization_from_B-scans} and \ref{sec:Demagnetization_Scaling}, as well as their theoretical value described by Eq.~\ref{eq:Demagnetization-factor_theory} correspond to the effective demagnetization factor.

\subsubsection{\label{sec:Demagnetization_from_B-scans}Determination of the demagnetization factor from field-scan experiments}

The value of the applied field at which $f_{\rm N}$ approaches zero ({\it i.e.} when the intermediate state disappears and a pure Meissner state sets in) corresponds to $B=(1-N) B_{\rm c}$ (Eq.~\ref{eq:Intermediate-state_Interval}). Linear fits of $f_{\rm N}(B_{\rm ap})$ data in the region of $0.1 \lesssim f_{\rm N} \lesssim 0.5$ (Fig.~\ref{fig:BnFn-Hscan}~b) result in $(1-N) B_{\rm c}=15.9(5)$, 12.4(4), 8.7(3), and 5.1(3)~mT for $T=0.25$, 1.75, 2.50, and 3.10~K, respectively. With the corresponding $B_{\rm c}$'s reported in Sec.~\ref{sec:B-scans}, the demagnetization factors are found to be $N(0.25{\rm ~K})=0.48(2)$, $N(1.75{\rm ~K})=0.46(2)$, $N(2.50{\rm ~K})=0.46(3)$, and $N(3.10{\rm ~K})=0.42(3)$. Note that the values of $N$ at $T=0.25$, 1.75, and 2.50~K are in perfect agreement with the theoretical value $N_{\rm th}=0.467$ of Eq.~\ref{eq:Demagnetization-factor_theory}. The small difference between $N(3.10{\rm ~K})$ and $N_{\rm th}$ is probably related to the absence of a pure Meissner state at $T=3.10$~K as reported above.

\subsubsection{\label{sec:Demagnetization_Scaling}Determination of the demagnetization factor from the scaled $f_{\rm N}(B_{\rm ap})$ and $f_{\rm N}(T)$ data}

\begin{figure}[htb]
\centering
\includegraphics[width=0.9\linewidth]{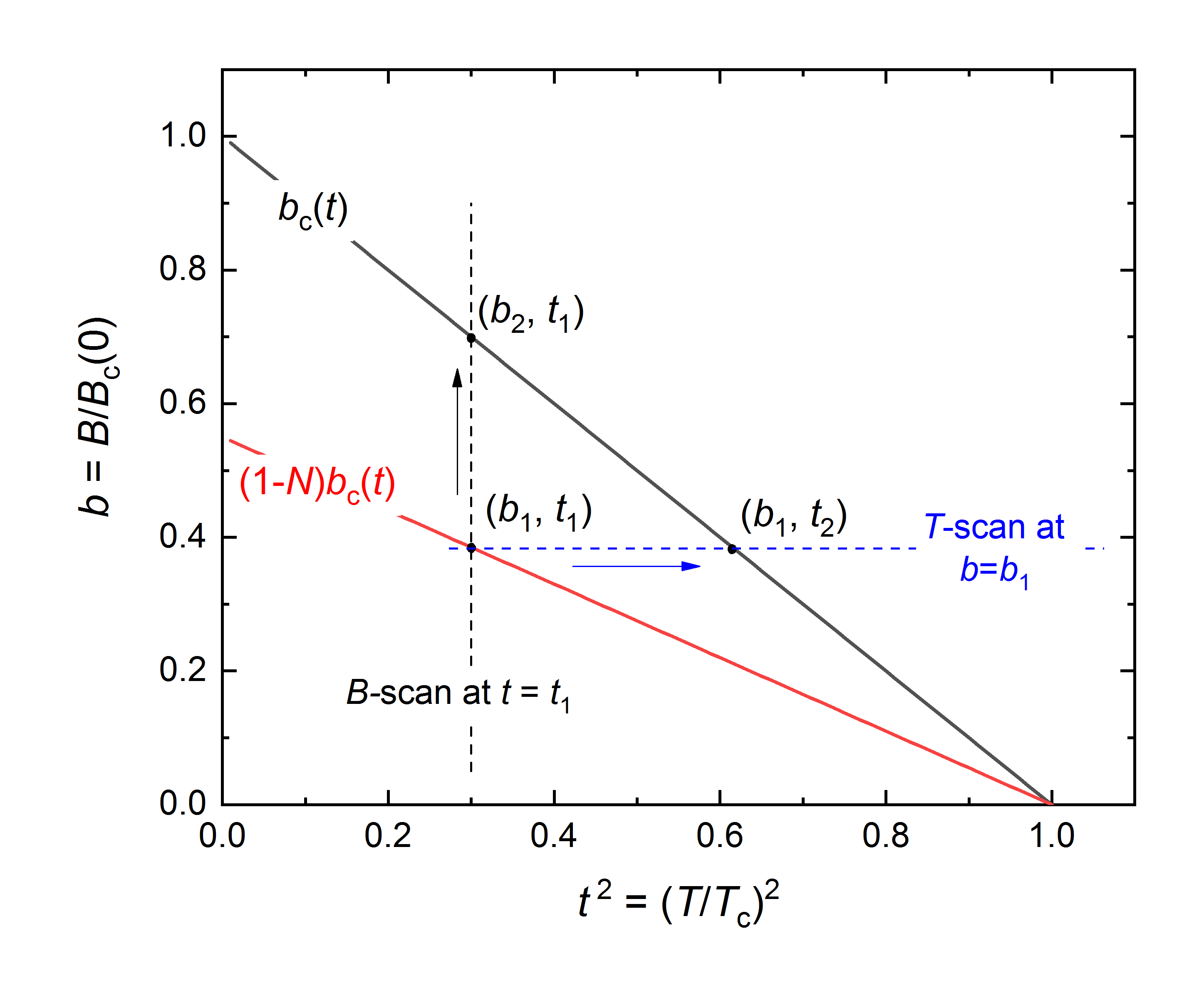}
\caption{The generic $B-T$ phase diagram of a type-I superconductor plotted in reduced field [$b=B_{\rm ap}/B_{\rm c}(T=0)]$) and reduced temperature [$t=T/T_{\rm c}(B_{\rm ap}=0)$] units. The reduced thermodynamic critical field $b_{\rm c}=B_{\rm c}(T)/B_{\rm c}(T=0)$ follows the parabolic dependence $b_{\rm c}(t)=1-t^2$. The field separating the pure Meissner state and the intermediate state is determined by $b_{\rm c}(1-N)$. The black and blue arrows represent the field-scan at $t=t_1$ and the temperature-scan at $b=b_1$. }
\label{fig:Scaling}
\end{figure}

The $f_{\rm N}(B_{\rm ap})$ and $f_{\rm N}(T)$ dependencies, as obtained in field-scan and temperature-scan measurements, can be scaled to single curves by using the procedure described in the following.
Figure~\ref{fig:Scaling} shows schematically the generic $B-T$ phase diagram of a type-I superconductor plotted in reduced field [$b=B_{\rm ap}/B_{\rm c}(T=0)]$) and reduced temperature[($t=T/T_{\rm c}(B_{\rm ap}=0)$] units. The reduced thermodynamic critical field $b_{\rm c}=B_{\rm c}(T)/B_{\rm c}(T=0)$ is assumed to be described as $b_{\rm c}(t)=1-t^2$. Note that in type-I superconductors $B_{\rm c}$ was found to follow closely the parabolic law:  $B_{\rm c}(T)=B_{\rm c}(0)[1-(T/T_{\rm c})^2]$.\cite{Tinkham_75, Poole_Book_2014} A field- and a temperature-scan are denoted by the black and the blue arrow, respectively. The normal state volume fraction $f_{\rm N}$ is equal to '0' at $(b_1,t_1)$ and increases to '1'  by approaching the $(b_2,t_1)$ or $(b_1,t_2)$ phase points. Assuming that $f_{\rm N}$ changes linearly between $b_1$ and $b_2$ in a field-scan and that it follows a $1-t^2$ behavior in a temperature-scan one gets:
\begin{equation}
f_{\rm N}(b,t=t_1)=\frac{b-b_1}{b_2-b_1} \nonumber
\end{equation}
and
\begin{equation}
f_{\rm N}(t,b=b_1)=\frac{t^2-t_1^2}{t_2^2-t_1^2}. \nonumber
\end{equation}
Taking into account that $b_2=b_c(t_1)$, $b_1=(1-N) b_c(t_1)$, and $t_2=t_c(b_1)$, simple mathematics gives:
\begin{equation}
f_{\rm N}(b,t=t_1)=\frac{b}{b_c(t_1)}\frac{1}{N}-\frac{1-N}{N}
 \label{eq:fn_reduced_b}
\end{equation}
and
\begin{equation}
f_{\rm N}(t,b=b_1)=\frac{1-t^2}{t_c(b_1)^2-1}\frac{1-N}{N}+\frac{1}{N}.
 \label{eq:fn_reduced_t}
\end{equation}

\begin{figure}[htb]
\centering
\includegraphics[width=0.85\linewidth]{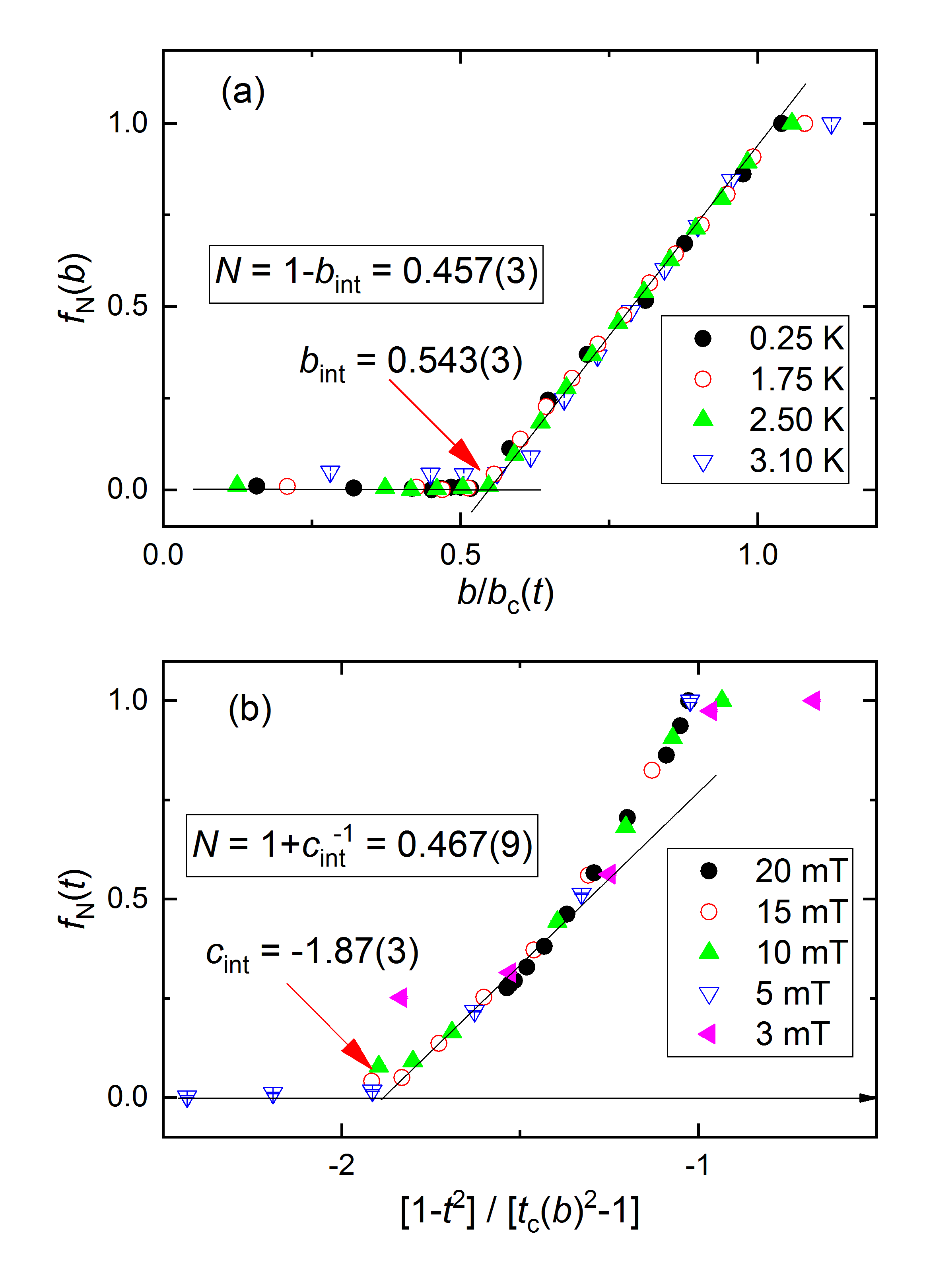}
\caption{(a) Dependence of the normal state domain's volume fraction $f_{\rm N}$ obtained in $B$-scan experiments on $b/b_{\rm c}(t)$. The intersection point $b_{\rm int}=0.543(3)$ determines the demagnetization factor $N=1-b_{\rm int}=0.457(3)$. (b) $f_{\rm N}$ from $T$-scan experiments as a function of $[1-t^2]/[t_c(b)^2-1]$. The intersection point $c_{\rm int}=-1.87(3)$ determines the demagnetization factor $N=1+c_{\rm int}^{-1}=0.467(9)$.}
\label{fig:Fn_scaled}
\end{figure}

Equation~\ref{eq:fn_reduced_b} implies that $f_{\rm N}(B_{\rm ap})$ dependencies measured at different temperatures (Fig.~\ref{fig:BnFn-Hscan}~b)  scale to a single curve by normalizing them to the corresponding $B_{\rm c}(T)$ values. This is indeed the case as is seen from the data presented in Fig.~\ref{fig:Fn_scaled}~a. The value of the demagnetization factor can be obtained from the intersection of the linear fit of $f_{\rm N}(b)$ data with the $f_{\rm N}=0$ line (Fig.~\ref{fig:Fn_scaled}~a) resulting in $b_{\rm int}=0.543(3)$. Following Eq.~\ref{eq:fn_reduced_b}, the demagnetization factor in field-scan experiments is found to be: $N=1-b_{\rm int}=0.457(3)$.

On the other hand the $f_{\rm N}(T)$ dependencies can be scaled by plotting them as a function of $[1-t^2]/[t_c(b)^2-1]$ (see Eq.~\ref{eq:fn_reduced_t} and Fig.~\ref{fig:Fn_scaled}~b). The value of the demagnetization factor obtained from the intersection point $c_{\rm int}=-1.87(3)$ results in: $N=1+c^{-1}=0.467(9)$. It should be noted here that, in contrast to  $f_{\rm N}(b)$ curves, the $f_{\rm N}(t)$ ones do not increase linearly in the region of $0< f_{\rm N} <1$ (Fig.~\ref{fig:Fn_scaled}~b). This suggests that our assumption of a $1-t^2$ behavior of $f_{\rm N}$ in $T$-scan experiments, which has been used to derive Eq.~\ref{eq:fn_reduced_t}, may not be fully correct.

\vspace{0.5cm}

The conclusions from the results presented in Sec.~\ref{sec:Demagnetization} are twofold: \\
(i) The values of the demagnetization factor $N$, as estimated from the $B_{\rm ap}$ and $T$ dependencies of the normal state domains volume fraction $f_{\rm N}$, are in good agreement with the value $N_{\rm th}=0.467$ based on the theoretical results of Ref.~\onlinecite{Prozorov_PRAppl_2018}.\\
(ii) The agreement between the theory and the experiment suggests that TF-$\mu$SR measurements can be used for studies of different materials in more complex geometries. A Sn probe or any other type-I superconductor of the same geometry as the sample under investigation, could be used as a reference for an experimental determination of the demagnetization factor.

\subsection{\label{sec:thermodynamic-field}Temperature dependence of the thermodynamic critical field}

\begin{figure}[htb]
\centering
\includegraphics[width=0.85\linewidth]{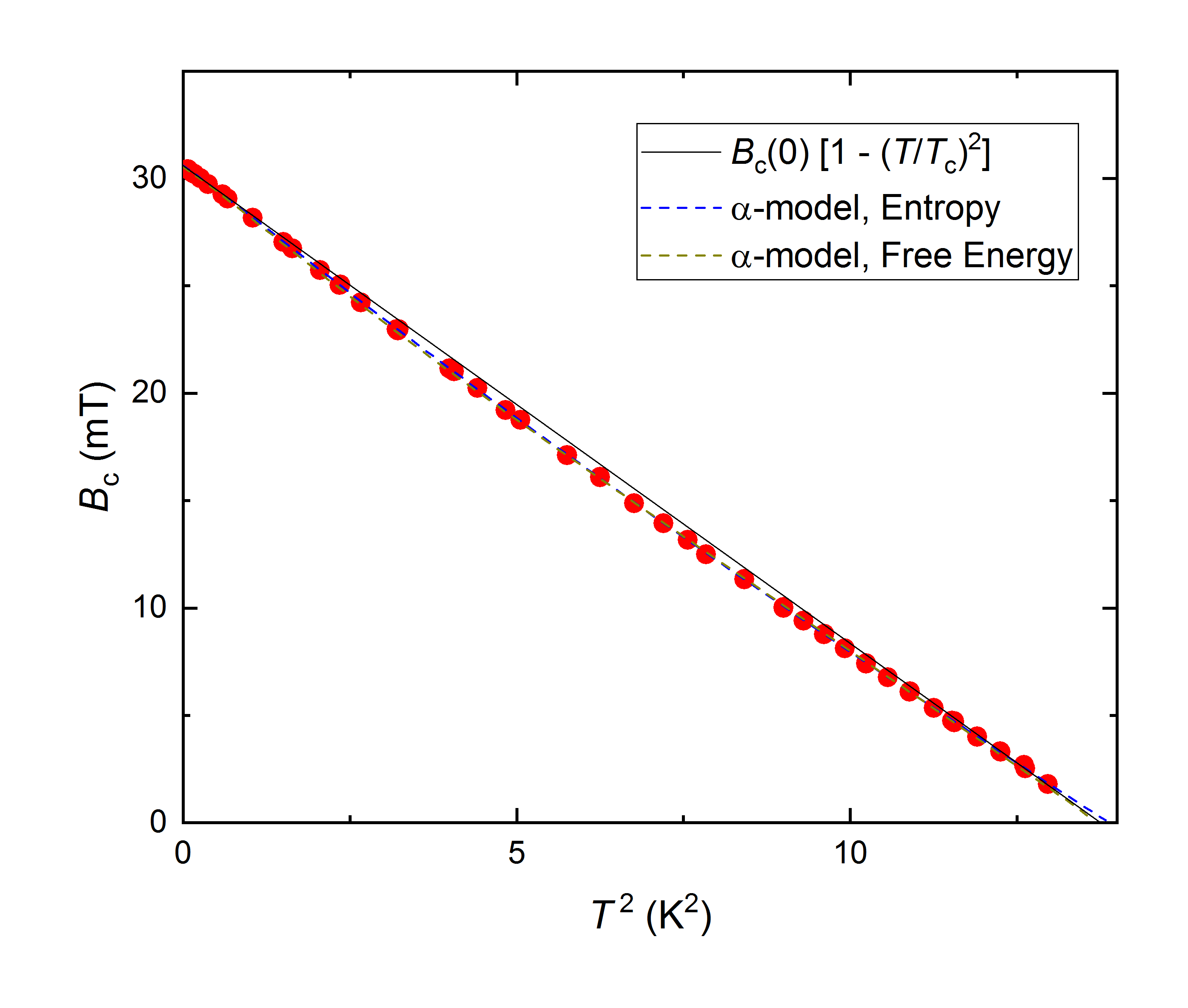}
\caption{Dependence of the thermodynamic critical field $B_{\rm c}$ on $T^2$ for $\beta-$Sn. The points obtained in $B$-scans (Sec.~\ref{sec:B-scans}), $T$-scans (Sec.~\ref{sec:T-scans}), and $B-T$-scans (Sec.~\ref{sec:B-T-scan}) experiments are included in the figure. The solid line corresponds to a parabolic $B_{\rm c}(T)$ behavior. Dashed lines are fits within the framework of the phenomenological $\alpha-$model by following the entropy and the free energy approaches (see text for details).}
\label{fig:Bc_full}
\end{figure}

Figure \ref{fig:Bc_full} shows the thermodynamic critical field $B_{\rm c}$ as a function of $T^2$. Points obtained in $B$-scans (Sec.~\ref{sec:B-scans}), $T$-scans (Sec.~\ref{sec:T-scans}), and $B-T$-scans (Sec.~\ref{sec:B-T-scan}) mesurements are plotted together. The experimental $B_{\rm c}(T^2)$ points follow rather well the parabolic behavior (black solid line), which is generally expected for type-I superconductors.\cite{Tinkham_75, Poole_Book_2014}
The deviation of $B_{\rm c}(T^2)$ from $B_{\rm c}(0)[1-(T/T_{\rm c})^2]$ was further investigated to estimate various thermodynamic quantities of the superconducting $\beta-$Sn within the framework of the phenomenological $\alpha-$model.

\subsubsection{\label{sec:alpha-model}$\alpha-$model}

Originally, the $\alpha$-model was adapted from the single-band Bardeen-Cooper-Schrieffer (BCS) theory of superconductivity in order to explain deviations of the temperature behavior of the electronic heat capacity and the thermodynamic critical field from the weak-coupled BCS prediction. The model assumes that the temperature dependence of the normalized superconducting energy gap:
\begin{equation}
\delta(T)=\frac{\Delta(T)}{\Delta}=\frac{\Delta_{\rm BCS}(T)}{\Delta_{\rm BCS}},
\label{eq:delta(t)}
\end{equation}
($\Delta$ is the zero-temperature value of the gap) is the same as in the BCS theory,\cite{Muehlschlegel_ZPhys_1959} and it is calculated for the BCS value $\alpha_{\rm BCS}=\Delta_{\rm BCS}/k_{\rm B}T_{\rm c}\simeq 1.764$ ($k_{\rm B}$ is Boltzmann's constant). On the other hand, to calculate the temperature evolution of the electronic free energy, the entropy, the heat capacity and the thermodynamic critical field, the $\alpha-$model allows $\alpha=\Delta/k_{\rm B} T_{\rm c}$ to be an adjustable parameter.

The single-band $\alpha-$model was originally developed by Padamasee {\it et al.}\cite{Padamsee_JLTP_1973} A detailed description of the single-band $\alpha-$model was recently given by Johnston in Ref.~\onlinecite{Johnston_SST_2013}. The two-band version of the $\alpha-$model is widely used to analyze the specific heat and the superfluid density data in compounds where more then one band are supposed to be involved in the superconducing mechanism.\cite{Bouquet_EPL_2001, Carrington_2003, Guritanu_PRB_2004, Prozorov_SST_2006, Khasanov_PRL_2007, Khasanov_JSNM_2008, Khasanov_PRB_2014, Khasanov_Arxiv_2018_1144}

\begin{figure}[htb]
\centering
\includegraphics[width=0.85\linewidth]{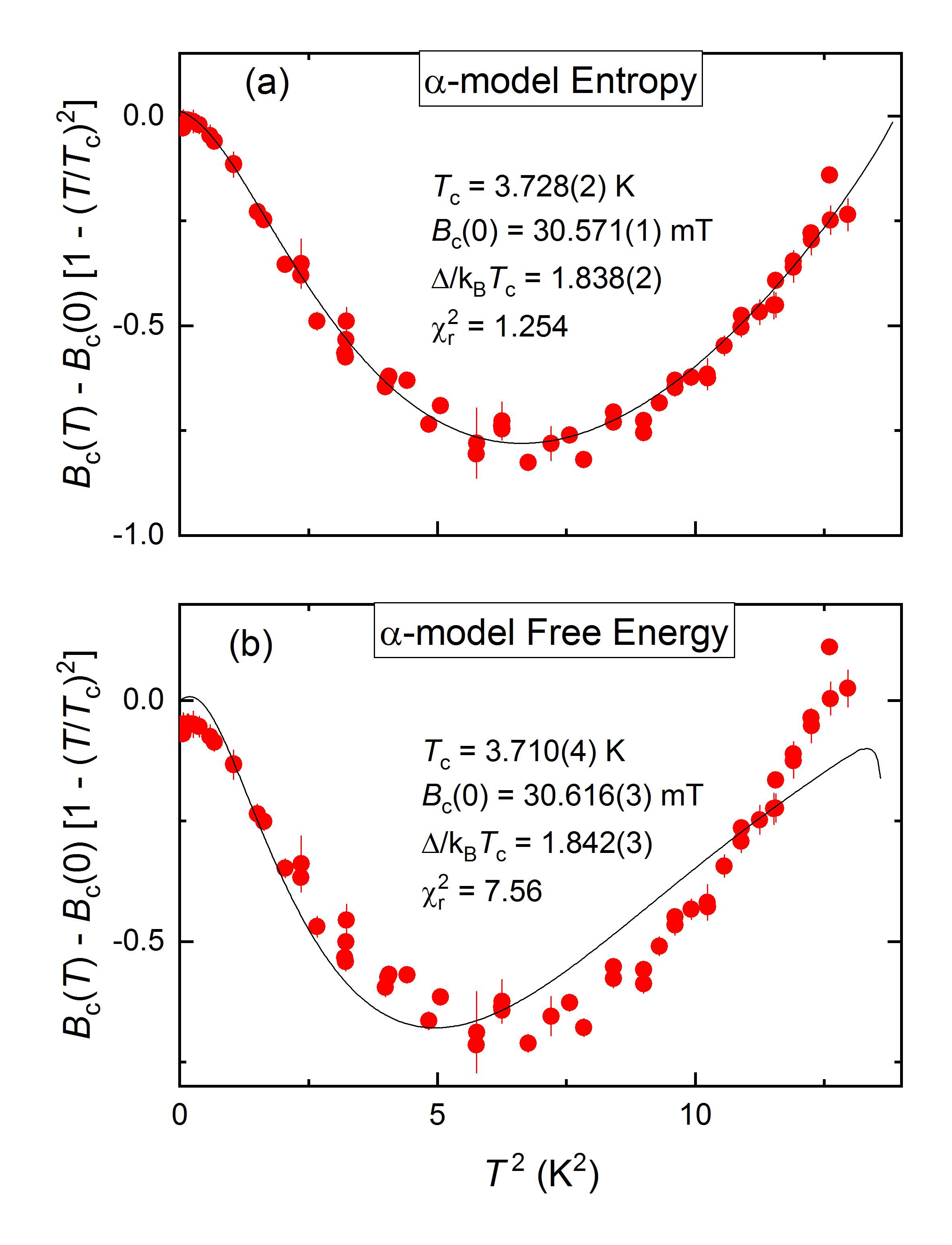}
\caption{(a) The deviation function $D(T)= B_{\rm c}(T)-B_{\rm c}(0)[1-(T/T_{\rm c})^2]$ obtained within the framework of the $\alpha-$model with an entropy approach. The points are the experimental data. The solid line is the fit of Eq.~\ref{eq:Bc_Entropy} to $B_{\rm c}(T)$ (Fig.~\ref{fig:Bc_full}). (b) The same as in panel (a) but for a free energy approach. The solid line is the fit of Eq.~\ref{eq:Bc_FreeEn} to $B_{\rm c}(T)$ data.}
\label{fig:alpha-deviations}
\end{figure}

The thermodynamic critical field, which is directly related to the condensation energy of the Cooper pairs $B_{\rm c}^2/8\pi$, can be obtained by using the difference in free energy as well as in entropy between normal and superconducting state (see Ref.~\onlinecite{Johnston_SST_2013}).

\subsubsection{\label{sec:alpha-model_free-energy}$B_{\rm c}(T)$: free energy approach}

Within the free energy picture:
\begin{equation}
\frac{B_{\rm c}^2(t)}{8 \pi}= F_{N}(t) - F_{\rm S}(t),
 \label{eq:Bc_FreeEn}
\end{equation}
with the normal state ($F_{\rm N}$) and the superconducting state ($F_{\rm S}$) free energy given by:\cite{Johnston_SST_2013}
\begin{equation}
\frac{F_{\rm N}(t)}{\gamma_{\rm e} T_{\rm c}^2}=-\frac{t^2}{2} \nonumber
\end{equation}
and
\begin{equation}
\frac{F_{\rm S}}{\gamma_e T_{\rm c}^2} = -\frac{3 \alpha^2}{\pi^2}
\left[  \frac{\delta(t)^2}{4} + \int_0^\infty f(\alpha,E,t) \frac{2\varepsilon^2 + \delta(t)^2}{E} d\varepsilon \right]. \nonumber
\end{equation}
Here, $f(\alpha,E,t)=[\exp(\alpha E/t) + 1]^{-1}$ is the Fermi function, $E=E(\varepsilon,\delta(t))=\sqrt{ \varepsilon^2 + \delta(t)^2 }$ the quasiparticle energy, and $\gamma_e$ is the normal state electronic specific heat coefficient. The temperature dependence of the normalized gap, tabulated by M\"uhlschlegel in Ref.~\onlinecite{Muehlschlegel_ZPhys_1959} can be parameterized as $\delta(t)=\tanh \{1.82[1.018(1/t-1)]^{0.51}\}$.\cite{Carrington_2003} Note that the Eq.~\ref{eq:Bc_FreeEn} and the forthcoming Eq.~\ref{eq:Bc_Entropy} are expressed in cgs units in analogy with Johnston.\cite{Johnston_SST_2013} The fit of Eq.~\ref{eq:Bc_FreeEn} to the $B_{\rm c}(T)$ data is shown by the dashed dark-yellow line in Fig.~\ref{fig:Bc_full}. The fit values for elemental $\beta-$Sn are: $T_{\rm c}=3.710(4)$~K, $B_{\rm c}(0)=30.616(3)$~mT and $\alpha=1.842(3)$.

\subsubsection{\label{sec:alpha-model_entropy}$B_{\rm c}(T)$: entropy approach}

The temperature evolution of $B_{\rm c}$ can be also determined from the difference between the normal state and the superconducting state entropies $S_{\rm N}-S_{\rm S}$ via:\cite{Padamsee_JLTP_1973, Johnston_SST_2013}
\begin{equation}
\frac{B_{\rm c}^2}{8\pi}=T_{\rm c} \int_t^1 \left[ S_{\rm N}(t') - S_{\rm S}(t') \right] dt'
  \label{eq:Bc_Entropy}
 \end{equation}
with
\begin{equation}
\frac{S_{\rm N}(t)}{\gamma_e T_c} = t \nonumber
\end{equation}
and
\begin{equation}
\frac{S_{\rm S}(t')}{\gamma_e T_{\rm c}} = \frac{6 \alpha^2}{\pi^2 t} \int_0^\infty f(\alpha, E,t)\left( E+ \frac{\varepsilon^2}{E} \right) d\varepsilon. \nonumber
\end{equation}

The fit of Eq.~\ref{eq:Bc_Entropy} to the $B_{\rm c}(T)$ data is shown by the dashed black line in Fig.~\ref{fig:Bc_full}.  In this case the following fit values are obtained: $T_{\rm c}=3.728(2)$~K, $B_{\rm c}(0)=30.571(1)$~mT and $\alpha=1.838(2)$.

\begin{table*}[htb]
\caption{\label{tab1} Experimentally obtained and calculated material parameters for $\beta-$Sn. $T_{\rm c}$ is the superconducting transition temperature,  $B_{\rm c}(0)$ is the zero-temperature value of the thermodynamic critical field, $\Delta$ is the zero-temperature value of the superconducting energy gap, $2\Delta/k_{\rm B}T_{\rm c}$ is the coupling strength, $\gamma_{\rm e}$ is the normal state electronic specific heat coefficient, and $\Delta C(T_{\rm c})/\gamma_e T_{\rm c}$ is the specific heat jump at $T_{\rm c}$.    }
\begin{tabular}{cccccccccc}
\hline
\hline
$T_{\rm c}$ &$B_{\rm c}(0)$&$\Delta$&$2\Delta/k_{\rm B}T_{\rm c}$&$\gamma_{\rm e}$&$\Delta C(T_{\rm c})/\gamma_e T_{\rm c}$& References\\
(K)& (mT) &(meV)&&(${\rm mJ}\;{\rm mol}{\rm K}^{-2}$)&& \\
\hline3.717(3)    & 30.578(6) & --& --& --&--& This work (experiment)\\
3.710(4)& 30.616(3)& 0.589(7)&3.684(6)& 1.713(3)& 1.555(4)&This work ($\alpha-$model, free energy)\\
3.728(2)& 30.571(2)& 0.590(6)&3.676(4)& 1.781(3)& 1.548(3)&This work ($\alpha-$model, entropy)\\
3.701--3.722&30.3--30.6 & 0.55--0.61 & 3.46--3.71&1.74--1.80&1.50--1.68& Refs.~ \onlinecite{Matthias_RMP_1963, Carbotte_RMP_1990, Padamsee_JLTP_1973, Finnemore_PR_1965, Bryant_PRL_1960, Corak_PRB_1956, Giaever_PR_1962, Richards_PR_1960, ONeal_PhysRev_1965, Douglass_PhysRev_1964, Walmsley_CanJPhys_1967, Townsend_Phys_Rev_1962} \\
\hline
\end{tabular}
\end{table*}

\subsubsection{\label{sec:alpha-model_comparison_entropy-free_energy}Comparison between the entropy and free energy approaches}

From Fig.~\ref{fig:Bc_full} it is not possible at first sight to establish which one of the two above discussed approaches better describes the $B_{\rm c}(T)$ behavior. A better comparison can be made by plotting the deviation of $B_{\rm c}(T)$ from the parabolic function: $D(T)= B_{\rm c}(T)-B_{\rm c}(0)[1-(T/T_{\rm c})^2]$, with the parameters $T_{\rm c}$ and $B_{\rm c}(0)$ obtained from the fits of Eqs.~\ref{eq:Bc_FreeEn} and \ref{eq:Bc_Entropy} to the experimental data.  Figure \ref{fig:alpha-deviations} shows the $D(T^2)$ deviation functions for both, entropy (panel a) and free energy (panel b) approaches. Obviously, the entropy expression gives a  much better agreement with the experimental data. The reduced $\chi^2_{\rm r}$ ({\it chi-square/number of degrees of freedom minus one}) is in this case $\chi^2_{\rm r}\simeq 1.254$ whereas for the free energy approach we obtain $\chi^2_{\rm r}\simeq 7.56$.

The disagreement between the two methods is quite surprising, since one would expect that both methods are consistent. Such a discrepancy was already noticed by Padamsee {\it et al.}\cite{Padamsee_JLTP_1973} in their original paper, where the phenomenological $\alpha-$model was introduced for the first time. The reason is that under the assumption of a BCS like temperature evolution of the superconducting energy gap, $\Delta(T)/\Delta\equiv\Delta_{\rm BCS}(T)/\Delta_{\rm BCS}$, Eqs.~\ref{eq:Bc_FreeEn} and \ref{eq:Bc_Entropy} are not equivalent to each other  for {\it any} $\alpha$ values except for $\alpha\equiv\alpha_{\rm BCS}=1.764$. Our numerical simulations of the $D(t)$ curves for various $\alpha$ values confirms this. Padamsee {\it et al.}\cite{Padamsee_JLTP_1973} have also shown that the analysis of the specific heat and the $B_{\rm c}(T)$ data of metallic Indium  within the framework of the free energy calculations result in $\alpha$ values which are more than $30\%$ different from each other.

Our $\mu$SR measurement on $\beta-$Sn provides, therefore, a further example of inconsistency between these two applications of $\alpha-$model and points towards the validity of the use of the entropy expression for the determination of the thermodynamical critical field. We should emphasize, however, that the detailed reason why the entropy approach gives better results than the free energy one is not clear and further theoretical work is necessary to understand this point.

\subsection{\label{sec:quantities_comparison}Comparison of physical quantities obtained from TF-$\mu$SR experiments with the literature data}

Table~\ref{tab1} summarizes the physical quantities obtained in the present study and compares them with those reported in the literature.
The 'experimental' values of $T_{\rm c}$ and $B_{\rm c}(0)$ were obtained from the intersection of linear fits of $B_{\rm c}(T^2)$ curve in the vicinity of $T\sim 0$ and $T\sim T_{\rm c}$ with the $T^2=0$ and $B_{\rm c}=0$ lines. The electronic specific heat $\gamma_{\rm e}$ was obtained from Eqs.~\ref{eq:Bc_FreeEn} and \ref{eq:Bc_Entropy} in the limit of $T\rightarrow 0$. The specific heat jump at the transition temperature $\Delta C(T_{\rm c})/\gamma_{\rm e} T_{\rm c}$, was calculated by using:\cite{Johnston_SST_2013}
\begin{equation}
\frac{ \Delta C(T_{\rm c})}{\gamma_{\rm e} T_{\rm c}}=-\left( \frac{3 \alpha^2}{2 \pi^2} \right) \left.\frac{d  \delta(t)^2}{d t} \right|_{t=1}
\simeq 1.426 \left( \frac{\alpha}{\alpha_{\rm BCS}} \right)^2.
\end{equation}

Table~\ref{tab1} shows that the physical quantities derived for elemental $\beta-$Sn in TF-$\mu$SR experiments are in agreement with the literature data. The only exception is $\gamma_{\rm e}=1.713(3)$, which was obtained by using the free energy approach. This again points to inadequacies of a 'free energy' based $\alpha-$model.

\section{\label{sec:Conclusions}Conclusions}

The type-I superconductivity of a high-quality cylindrically shaped $\beta-$Sn sample was studied by means of the transverse-field muon-spin rotation/relaxation technique. In the intermediate state, {\it i.e.} when the type-I superconducting sample with non-zero demagnetization factor $N$ is separated into the normal state and the Meissner state domains, the $\mu$SR technique allows to determine with very high precision the value of the thermodynamic critical field $B_{\rm c}$, as well as the relative sample volumes occupied by various types of domains. Due to the microscopic nature of the technique the $B_{\rm c}$ values are determined {\it directly} via measurements of the internal field inside the normal state domains. No assumptions or introduction of any type of measurement analysis are needed.

The main results of the paper are summarized as follows: \\
(i) The full $B-T$ phase diagram of a cylindrical $\beta-$Sn sample was reconstructed. The transition temperature $T_{\rm c}=3.717(3)$~mT and the zero-temperature value of the thermodynamic critical filed $B_{\rm c}(0)=30.578(6)$~mT are found to be in full accordance with the literature data. \\
(ii) Measurements at various $(B,T)$ points within the phase diagram of $\beta-$Sn reveal that the local field inside the normal state domains ($B_{\rm N}$) is independent, within our experimental accuracy, on the relative sample volumes occupied by the normal state and superconducting domains, thus confirming the calculations presented in Refs.~\onlinecite{Tinkham_75, Egorov_PRB_2001, deGennes_Book_1966}. \\
(iii) Magnetic history effects caused by different paths in the $B-T$ phase space do not lead to measurable changes of the magnetic field distributions probed by means of $\mu$SR. This implies that even if the distribution and/or the shape of the domains are history dependent,\cite{Prozorov_PRL_2007, Prozorov_NatPhys_2008, Prozorov_JPCS_2009} the internal field inside the normal state domains, as well as the relative sample
volumes occupied by the normal state and the Meissner state domains remain unaltered. \\
(iv)  The values of the demagnetization factor $N$, as estimated from the $B_{\rm ap}$ and $T$ dependencies of the normal state domain's volume fraction $f_{\rm N}$, are in good agreement with the theoretical values $N_{\rm th}=0.467$  of Ref.~\onlinecite{Prozorov_PRAppl_2018}. The agreement between the theory and the experiment suggests that TF-$\mu$SR measurements can be used for studies of different materials in more complex geometries. A Sn probe, or any other type-I superconductor, of same geometry as a sample under investigation, can be used as a reference for the experimental determination of demagnetization factors.\\
(v) Analysis of the $B_{\rm c}(T)$ dependence within the framework of phenomenological $\alpha-$model allows to obtain the value of the superconducting energy gap $\Delta=0.59(1)$~meV, the electronic specific heat $\gamma_e=1.781(3)$~${\rm mJ}/{\rm mol}\; {\rm K}^2$ and the jump in the heat capacity ${\Delta C(T_c)}/{\gamma T_{\rm c}}=1.55(2)$. All these quantities are found in good agreement with the literature values. \\
(vi) Analysis of the experimental data by means of the $\alpha-$model within a 'free energy' and 'entropy' scenario reveals that the 'free energy' approach does not describe satisfyingly  the experimental data. This confirms the conclusion of Ref.~\onlinecite{Padamsee_JLTP_1973} about inconsistency of these two types of $\alpha-$model application and points to the validity of the 'entropy' approach.

\vspace{0.5cm}

\begin{acknowledgments}
The authors would like to thank Hans-Henning Klau{\ss} for helpful discussions.

The experiment was performed at the Swiss Muon Source (S$\mu$S, PSI Villigen, Switzerland). The work was advertised and will be available as an experiment at the Advanced Physics Laboratory of ETH, Zurich, Switzerland.
\end{acknowledgments}

\end{document}